\documentclass[12pt]{article}
\usepackage{amsfonts}
\usepackage{amsmath}
\usepackage{xcolor}
\usepackage{graphicx}
\usepackage{caption}
\usepackage{subcaption}
\usepackage{cite}
\usepackage{epstopdf}
\usepackage[section]{placeins}
\newcommand{\be}{\begin{equation}}
\newcommand{\ba}{\begin{eqnarray}}
\newcommand{\ee}{\end{equation}}
\newcommand{\ea}{\end{eqnarray}}

\newcommand{\sech} { {\rm sech}}

\begin{document}

\title{A Class of Exactly Solvable Real and Complex $PT$ Symmetric 
Reflectionless  Potentials}

\author{Suman Banerjee$^{a}$\footnote{e-mail address: suman.raghunathpur@gmail.com (S.B)}, Rajesh Kumar Yadav$^{a}$\footnote{e-mail address: rajeshastrophysics@gmail.com (R.K.Y)},  Avinash Khare$^{b}$\footnote {e-mail address: avinashkhare45@gmail.com (A.K)} and \\ Bhabani Prasad Mandal$^{c}$\footnote{e-mail address:bhabani.mandal@gmail.com(BPM)}}
\maketitle
{$~^a$Department of Physics, Sido Kanhu Murmu University, Dumka-814110, India.\\
$~^b$Department of Physics, Savitribai Phule Pune University, Pune-411007, India.\\
$~^c$Department of Physics, Banaras Hindu University, Varanasi-221005, India.}

\begin{abstract}
We consider the question of the number of exactly solvable complex 
but PT-invariant reflectionless potentials with $N$ bound states. By 
carefully considering the $X_m$ rationally extended reflectionless potentials, 
we argue that the total number of exactly solvable complex PT-invariant 
reflectionless potentials are $2[(2N-1)m+N]$.
\end{abstract}

PACS numbers: 11.30.Pb, 03.65.Ge, 03.65.Nk, 

\section{Introduction}
The reflectionless potentials, also known as transparent potentials or black 
potentials, are of great importance in physics and engineering.  
In view of the numerous applications of the reflectionless potentials, it is
very important to search for new reflectionless potentials. While it is well
known that there are $N$ continuous parameter families of exactly solvable 
real reflectionless potentials, to the best of our knowledge, the question 
of the complex PT-invariant exactly solvable reflectionless potentials has
not been addresses in the literature.

In the last two decades, after the discovery of the $PT$ (combined parity ($P$)
and time reversal ($T$)) symmetric non-hermitian
systems \cite{bender_98, bender_07, AM_10}, it has been shown that there are 
 non-hermitian complex PT-invariant potentials which are also 
reflectionless \cite{rlpt_05,rlpt_11,rlpt_14,rlpt_2011,rlpt_13}.
After the discovery of the $X_m$ exceptional orthogonal polynomials (EOPs) 
\cite{eopm1,eopm2,eopm3}, a group of new (also known as rationally extended)
hermitian as well as $PT$ symmetric non-hermitian 
potentials have been constructed \cite{que,bqr,os,mi_os,op1,op2,op3,midya,yg15,
rkyd,scatt1,scatt2,scatt3,n16,ramos,nk17,nk18,bbp,rkmany,satish} with their solutions in terms of these 
EOPs. It is then natural to enquire how many distinct reflectionless potentials 
with $N$ bound states can be constructed using both PT symmetry and $X_m$
EOPs. This is the task that we have undertaken in this paper.

Just to set the notation, we first consider a real exactly solvable 
reflectionless potential with $N$ bound states and using the method of 
supersymmetric quantum mechanics \cite{cks,ks,pursey,amp} explicitly obtain one
continuous ($\lambda$) parameter family of reflectionless potentials 
including the corresponding reflectionless Pursey and Abraham Moses (AM) 
potentials with $N-1$ bound states. This can be generalized and one can 
obtain $N$ continuous parameter families of real reflectionless potentials with
$N$ bound states.

We then consider the case of the non-hermitian $PT$ symmetric Scarf-II 
reflectionless potentials with $N$ bound states and discuss the role of 
the parametric symmetry. Finally, we consider the rationally extended 
complex PT-symmetric scarf-II potential whose eigenfunctions 
are written in the form of $X_m$ Jacobi EOPs with $m=1,2,3..$, and show that 
these extended potentials are also reflectionless and combining all these
factors we have in all $2[(2N-1)m+N]$ number of complex PT-symmetric 
reflectionless potentials with $N$ bound states. 

The organization of this manuscript is as follows: In Sec. $2$, we briefly 
discuss the formalism of supersymmetric quantum mechanics relevant to this 
paper \cite{cks} and explicitly obtain one continuous parameter family
of real reflectionless potentials with $N$ bound states. In Sec. $3$, we
consider the complex PT-invariant Scarf-II potential with $N$ bound states and
obtain conditions underwhich it is reflectionless. We also discuss the role of
parametric symmetry in counting the number of reflectionless PT-symmetric
complex potentials. We then construct the
corresponding $X_m$ family of complex PT-invariant reflectionless potentials 
with $N$ bound states and argue that the total number of complex PT-invariant 
reflectionless potentials is $2[(2N-1)m +N]$.
In Sec. $4$, we consider the case of $N=3$ explicitly. First we give  
 explicit expression as well as suitable plots for the one parameter family of 
 real reflectionless potentials
 and the corresponding eigenfunctions with three bound states. 
 We then give explicit expression for the real and imaginary parts of the
 complex PT-invariant potentials and their eigenfunctions in the case of three
 bound states. Finally, in Sec. $5$, we summarize our results.

 \section{Formalism}

In this section, we set the basic notations of supersymmetric quantum mechanics
(SQM) as relevant to the present discussion. We then discuss the case of the
real reflectionless potential with $N$ bound states and obtain one 
continuous parameter family of strictly isospectral reflectionless potentials 
with $N$ bound states. This can be generalized \cite{cks} and one can
obtain $N$ continuous parameter family of reflectioless real potentials with
$N$ bound states.

\subsection{Basic Results of Supersymmetric Quantum Mechanics}

Consider a  Hamiltonian 
\be\label{1}
H^{(1)}(x)=-\frac{d^2}{dx^2}+V^{(1)}(x), \quad (\hbar=2m=1)
\ee
 with ground state energy $E_{0}^{(1)} = 0$. One can then factorize $H^{(1)}$ 
in terms of the operators $A$ and $A^{\dagger}$ as 
\be\label{2}
H^{(1)}(x)=A^{\dagger} A 
\ee 
with
\be\label{3}
A=\frac{d}{dx}+W(x) \quad \mbox{and} \quad A^{\dagger}=-\frac{d}{dx}+W(x),
\ee
where 
\be\label{sup}
W(x) =-\frac{d}{dx}[\ln\psi^{(1)}_{0}(x)]
\ee 
is the superpotential, which determines the two partner potentials 
\ba\label{pp}
V^{(1)}(x)=W^2(x)-W'(x) \quad \mbox{and} \quad V^{(2)}(x)=W^2(x)+W'(x).
\ea
The eigenvalues and the eigenfunctions of these two potentials 
(when the SUSY is unbroken) are related by 
\be\label{ev}
E^{(1)}_{n+1}=E^{(2)}_n \qquad  E^{(1)}_{0}=0,
\ee
and 
\be\label{pwf}
\psi^{(2)}_{n}(x)=\frac{1}{[E^{(2)}_n]^{1/2}}A\psi^{(1)}_{n+1} \qquad 
\psi^{(1)}_{n+1}(x)=\frac{1}{[E^{(2)}_n]^{1/2}}A^{\dagger}\psi^{(2)}_{n}
\ee
respectively. For the one dimensional case, the transmission $(T^{(1,2)}(k))$ 
and reflection $(R^{(1,2)}(k))$ amplitudes 
for the partner potentials $V^{(1,2)}(x)$ are related by
\be\label{r}
R^{(1)}(k)=\bigg(\frac{W_{-}+ik}{W_{-}-ik}\bigg)R^{(2)}(k)
\ee
and 
\be\label{t}
T^{(1)}(k)=\bigg(\frac{W_{+}-ik'}{W_{-}-ik}\bigg)T^{(2)}(k)
\ee
where 
\be\label{kk}
k=(E -W^2_{-})^{\frac{1}{2}}\quad \mbox{and}\quad k'=(E -W^2_{+})^{\frac{1}{2}}
\ee
with      
\be\label{scatt}
W_{\pm} =W(x\rightarrow \pm \infty).
\ee

The one-parameter family of potentials $\hat{V}^{(1)}(\lambda,x)$ which are
strictly isospectral to the given potential $V^{(1)}(x)$ are given by 
\be\label{isop}
\hat{V}^{(1)}(\lambda,x) =V^{(1)}(x)-2\frac{d^2}{dx^2}\ln(I(x)+\lambda)\,,
\ee    
where the integral $I(x)$  in term of the normalized ground state wavefunction 
is given by
\be\label{intim}
I(x)=\int^{x}_{-\infty}[\psi^{(1)}_{0}]^2(x)\, dx
\ee
and $\lambda$ is a constant which is either $>0$ or $<-1$. 
The corresponding superpotential $\hat{W}(x)$ 
with the same SUSY partner potential $V^{(2)}(x)$ is given by 
\be\label{spiso}
\hat{W}(x)=W(x)+\frac{d}{dx}\ln[I(x)+\lambda].
\ee 
The associated normalized ground state wavefunctions to the potential 
$\hat{V}^{(1)}(\lambda,x)$ is given by 
\be\label{gswf}
\hat{\psi}^{(1)}_{0}(\lambda,x)=\frac{\sqrt{\lambda (1+\lambda )}
\psi ^{(1)}_{0}(x)}{[I(x)+\lambda]},
\ee
while the normalized excited-state ($n=1,2,3...$) eigenfunctions are given by
\be\label{extwf}
\hat{\psi}^{(1)}_{n+1}(\lambda,x)=\psi^{(1)}_{n+1}(x)
+\frac{1}{E^{(1)}_{n+1}}\bigg(\frac{I'(x)}{I(x)+\lambda}\bigg)
\bigg( \frac{d}{dx}+W_(x)\bigg)\psi^{(1)}_{n+1}(x).
\ee

\subsubsection{Pursey potential}
 
The superpotential for this case is defined by putting $\lambda=0$ in Eq. (\ref{spiso})    
\be\label{ps}
W^{[P]}(x) = W(x)+\frac{d}{dx}\ln I(x).
\ee
and the potential (\ref{isop}) becomes
\be\label{pur}
V^{[P]}(x)=\hat{V}^{(1)}(\lambda = 0,x) =V^{(1)}(x)-2\frac{d^2}{dx^2}\ln I(x)\,,
\ee
while the corresponding eigenvalues are
\be\label{eig}
E^{[P]}_n=E^{(2)}_n = E^{(1)}_{n+1}\,,~~n = 0,1,2... \,.
\ee 
The reflection and transmission amplitudes for this case are 
\be\label{refl}
R^{[P]}(k)=\bigg(\frac{W_{-}-ik}{W_{-}+ik}\bigg)^2 R^{(1)}(k)
\ee
\be\label{tamp}
T^{[P]}(k)=-\bigg(\frac{W_{-}-ik}{W_{-}+ik}\bigg)T^{(1)}(k)\,.
\ee

\subsubsection{Abraham-Moses potential} 

In this case the superpotential and the potential ($\lambda=-1$) are given by 
\be\label{psup}
W^{[AM]}(x)=W(x)+\frac{d}{dx}\ln (I(x)-1),
\ee
and 
\be\label{am}
V^{[AM]}(x)=\hat{V}^{(1)}(\lambda = -1,x)=V^{(1)}(x)-2\frac{d^2}{dx^2}\ln (I(x)-1).
\ee
The eigenvalues are identical to the Pursey potential and are given by 
Eq. (\ref{eig}). 
The reflection and transmission amplitudes for this case are
\be\label{ram}
R^{[AM]}(k)=R^{(1)}(k)
\ee
\be\label{tam}
T^{[AM]}(k)=-\bigg(\frac{W_{+}+ik'}{W_{+}-ik'}\bigg)T^{(1)}(k)\,.
\ee

\subsection {Real potentials with $N$-bound states}

We consider the well known example of real reflectionless potential 
 \be\label{real_pot}
V^{(1)}(x)=-N(N+1)\sech^{2}(x); \quad -\infty\le x \le \infty,
\ee 
for any positive integer $N>0$. The solutions of the time-independent 
one-dimensional schr\"odinger equation corresponding to this potential 
are  well known \cite{cks} and given as

 
 

\be\label{ref1} 
\psi^{(1)}_{n} (x)= C_{n}^{(N)}\sech^{N}(x) P^{(-N-\frac{1}{2},-N-\frac{1}{2})}_n(i\sinh(x)),
\ee
with the energy eigenvalues
\be\label{fgu} 
E^{(1)}_{n}=-(N-n)^2, \quad n=0,1,2...n_{max}< N
\ee
and the normalization constant
\be\label{rnc}
C_{n}^{(N)}=2^N\bigg[\frac{n!(N-n)[\Gamma(N-n+\frac{1}{2})]^2}
{\Gamma(2N-n+1)\pi}\bigg]^{1/2}.
\ee 
Here $P^{(-N-\frac{1}{2},-N-\frac{1}{2})}_n(i\sinh(x))$ is the Jacobi 
polynomial. 
The corresponding reflection amplitude is zero at all positive energies while
the transmission amplitude is given by
\be\label{rtrans}
T^{(1)}(k)=\frac{\Gamma(-N-ik)\Gamma(N-ik+1)}{\Gamma(1-ik)\Gamma(-ik)},
\ee
with $k^2 = E^{(1)}_n$ while the transmission probability $|T^{(1)}(k)|^2 =1$.

\subsubsection{One-parameter family of reflectionless potentials}

It is straight forward to obtain the one continuous parameter family of 
reflectionless potentials by using Eqs. (\ref{isop}),
(\ref{intim}) and (\ref{real_pot}) and we obtain
\be\label{9h}
\hat{V}^{(1)}(\lambda,x) = -N(N+1)\sech^2(x) -2\frac{d^2}{dx^2}
\ln(I(x)+\lambda)\,,
\ee    
where
\be\label{9j}
I(x) = [C_{0}^{N}]^{2} \int_{-\infty}^{x} \sech^{2N}(y) \, dy\,,
\ee
and $\lambda > 0$ or $\lambda < -1$. Further, it is straight forward to
obtain the corresponding reflectionless Pursey or AM reflectionless 
potentials with $N-1$ bound states using Eqs. (\ref{pur}) and (\ref{am}).

This procedure can be iterated $N$ times to find $N$ continuous parameter
family of strictly isospectral reflectionless potentials with $N$ bound
states \cite{cks}.

\section{$PT$ Symmetric Complex Reflectionless Potentials}

Apart from the $N$ continuous parameter families of real reflectionless 
potentials with $N$ bound states, it turns out that there are a vast number of
complex $PT$ symmetric reflectionless  potentials with $N$ bound states which
we discuss in this section by starting from the well known complex PT-invariant
Scarf-II potential.

\subsection{$PT$ symmetric complex Scarf-II potential}

The complex $PT$ symmetric Scarf-II potential giving entirely real spectrum is 
well-known \cite{pt_02,rtscarf} and given by
\be\label{spot1}
V^{(1)}(x,a,b)=-[b^2+a(a+1)]\sech^2 (x)+ib(2a+1)\sech (x)\tanh (x); -\infty<x<\infty.
\ee
The corresponding bound state energy eigenvalues and the eigenfunctions 
respectively are 
\be\label{a}
E^{(1)}_n=-(a-n)^2, \quad n=0,1,2....n_{max}<a, 
\ee
and 
\be\label{swf1}
\psi^{(1)}_n(x,a,b)=C^{(a,b)}_n (\sech x)^a \exp (-ib \tan^{-1}(\sinh x))P^{(\alpha,\beta)}_n(i\sinh x),
\ee
 with $\alpha=b-a-\frac{1}{2}$ and $\beta=-b-a-\frac{1}{2}$.
 
The transmission  and the reflection amplitudes of this potential are also
well known \cite{rtscarf} and are given by
\be\label{scatt1}
T^{(1)}_{scarf}(k,a,b)=\frac{\Gamma(-a-ik) \Gamma(1+a-ik) 
\Gamma(\frac{1}{2}-b-ik) \Gamma(\frac{1}{2}+b-ik)}{\Gamma(-ik)
\Gamma(1+ik) (\Gamma (\frac{1}{2}-ik))^2},
\ee
and
\be\label{scatt2}
 R^{(1)}_{scarf}(k,a,b) = T^{(1)}_{scarf}(k,a,b)\times i\bigg[\frac{\cos \pi a \sin \pi b}
 {\cosh \pi k}+ \frac{\sin \pi a \cos \pi b}{\sinh \pi k}\bigg]
\ee
respectively, where $k^2 = E^{(1)}_n$.

\subsubsection{Parametric Symmetry}

This potential (\ref{spot1}) is invariant under the parametric 
transformation $b\leftrightarrow a+\frac{1}{2}$, however the corresponding 
eigenvalues and eigenfunctions are different \cite{parascarf} i.e., 
\be
V^{(1,p)}(x,a,b)= V^{(1)}(x,a\rightarrow b-\frac{1}{2},b\rightarrow a+\frac{1}{2}) 
= V^{(1)}(x,a,b)
\ee
but
\be\label{engp1}
E^{(1,p)}_n=-(b-n-\frac{1}{2})^2;\qquad n=0,1,2....n_{\mbox{max}}<b-\frac{1}{2},
\ee
and  
\be\label{swf2}
\psi^{(1,p)}_n(x,a,b)=\psi^{(1)}_n(x,a\rightarrow b-\frac{1}{2},b\rightarrow a+\frac{1}{2}).
\ee
Here $p$ denotes the quantities obtained after parametric transformation.
However, it is easy to check that the corresponding reflection and 
transmission amplitudes as given by Eqs. (\ref{scatt1}) and (\ref{scatt2})
are invariant under the parametric transformation 
$b\leftrightarrow a+\frac{1}{2}$.

Since the potential (\ref{spot1}) has two different sets of  
eigenvalues and eigenfunctions $\psi^{(1)}_n(x,a,b)$ and $\psi^{(1,p)}_n(x,a,b)$, 
hence there are two different superpotentials corresponding to the same
potential (\ref{spot1}) and are given by
\be
W(x,a,b)= a\tanh(x)+ib\sech(x)\,,
\ee
and
\be
W^{(p)}(x,a,b)= (b-\frac{1}{2})\tanh(x)+i(a+\frac{1}{2})\sech(x)\,.
\ee
This in turn gives two different partner potentials 
\ba\label{spotp1}
V^{(2)}(x,a,b)&=&(W(x,a,b))^2+ W(x,a,b)'\nonumber\\
&=&-(b^2+a(a-1))\sech^2 (x)+ib(2a-1)\sech (x)\tanh (x)
\ea
and 
\ba\label{spotp2}
V^{(2,p)}(x,a,b)&=&(W^{(p)}(x,a,b))^2+W^{(p)}(x,a,b)'\nonumber\\
&=&-((b-1)^2+a(a+1))\sech^2 (x)+i(b-1)(2a+1)\sech (x)\tanh (x)\nonumber\\
\ea
respectively. The first partner potential (\ref{spotp1}) is SI under translation of parameter $a\rightarrow a-1$, whereas 
the second one (\ref{spotp2}) is SI under translation of another parameter $b\rightarrow b-1$. 
The reflection and transmission amplitudes of these two partner potentials are 
related to those of the potential (\ref{spot1}) by
\ba\label{rscarf}
{R}^{(2)}_{scarf}(k,a,b)&=&\bigg(\frac{W_{-}-ik}
{W_{-}+ik}\bigg) R^{(1)}_{scarf}(k,a,b )\nonumber\\
&=&\bigg(\frac{a+ik}{a-ik}\bigg) R^{(1)}_{scarf}(k,a,b)\,,
\ea
\be\label{tscarf}
{T}^{(2)}_{scarf}(k,a,b)=-\bigg(\frac{a+ik}{a-ik}\bigg){T}^{(1)}_{scarf}(k,a,b)\,.
\ee
In the parametric case, the reflection and transmission amplitudes are invariant i.e., 
\ba
{R}^{(1,p)}_{scarf}(k,a,b)&=&{R}^{(1)}_{scarf}(k,a\rightarrow b-\frac{1}{2},b\rightarrow a+\frac{1}{2})\nonumber\\
{T}^{(1,p)}_{scarf}(k,a,b)&=&{T}^{(1)}_{scarf}(k,a\rightarrow b-\frac{1}{2},b\rightarrow a+\frac{1}{2})\nonumber\\
\ea
and for the corresponding partner potentials these are related as
\ba\label{rscarfp}
{R}^{(2,p)}_{scarf}(k,a,b)&=&\bigg(\frac{W^{(p)}_{-}-ik}
{W^{(p)}_{-}+ik}\bigg) R^{(1,p)}(k,a,b )\nonumber\\
&=&\bigg(\frac{b-\frac{1}{2}+ik}{b-\frac{1}{2}-ik}\bigg) R^{(1,p)}_{scarf}(k,a,b)\,,
\ea
\be\label{tscarf}
{T}^{(2,p)}_{scarf}(k,a,b)=-\bigg(\frac{b-\frac{1}{2}+ik}{b-\frac{1}{2}-ik}\bigg){T}^{(1,p)}_{scarf}(k,a,b)\,
\ee
respectively.
\subsubsection{Conditions for Reflectionless potentials}

From the Eq. (\ref{scatt2}), it follows that all the potential $V^{(1)}(x,a,b)$
and hence $V^{(2)}(x,a,b)$ and $V^{(2,p)}(x,a,b)$ are reflectionless when 
the potential parameters $a$ and $b$ are either both integers or both half 
integers. Remarkably, using parametric symmetry it turns out that there are 
in fact $2N$ distinct complex $PT$-invariant
reflectionless potentials all of which hold $N$ bound states. Out of these 
$N$ complex PT-invariant reflectionless potentials have half-integral values 
of $a$ and $b$ while the remaining $N$ complex PT-invariant potentials have
integral values of $a$ and $b$ which we now list one by one. 
 
{\bf {Case (A): If $a$ and $b$ both half integers}} 
\be\label{hint}
[a,b]=[(2N-1)/2,1/2],[(2N-3)/2, 3/2],...,[3/2,(2N-3)/2],[1/2,(2N-1)/2]\,.
\ee 
On using the fact that the eigenvalues of the complex Scarf-II potential are
given by Eq. (\ref{a}) while those of the corresponding parametric case are
given by Eq. (\ref{engp1}), one can immediately figure out about how many 
eigenvalues are coming from the normal scarf-II and how many from the 
parametric case. For example, while in the case $[a=(2N-1)/2,b=1/2]$, all the
$N$ eigenvalues are from normal scarf-II, in the case $[a=(2N-3)/2,b = 3/2]$, 
while $N-1$ eigenvalues are coming from normal Scarf-II, one 
eigenvalue is coming from the parametric case.

{\bf {Case (B): If $a$ and $b$ both integers}} 
\be\label{int}
[a,b]=[N-1,1],[N-2, 2],...,[1,N-1],[0,N]\,.
\ee
In the case $[a=0,b=N]$, while all the $N$ eigenvalues are from the 
parametric case, in the case of $[a = 1,b = N-1]$ we have $1$ eigenvalue
from the normal Scarf-II while $N-1$ eigenvalues are coming from the 
parametric case.

\subsection{Rationally Extended $PT$ Symmetric Complex Potential}

The $PT$ symmetric complex Scarf-II potential $V^{(1)}(x,a,b)$ (given by 
Eq. (\ref{spot1})) has been extended rationally \cite {midya} in terms of 
classical Jacobi polynomials $P^{(\alpha,\beta)}_m (z)$ for any positive 
integers of $m\ge0$ given by
\ba\label{scarfextdpot}
V^{(1)}_{m,ext}(x,a,b)&=&V^{(1)}(x,a,b)+2m(2b-m+1)+(2b-m+1)\nonumber\\
&\times &[(-2a-1)+(2b+1)i\sinh x]\bigg(\frac{P^{(-\alpha,\beta)}_{m-1}(i\sinh x)}{P^{(-\alpha-1,\beta-1)}_{m}(i\sinh x)}\bigg)\nonumber\\
&-&\frac{(2b-m+1)^2\cosh^2x}{2}\bigg (\frac{P^{(-\alpha,\beta )}_{m-1}(i\sinh x)}
{P^{(-\alpha-1,\beta-1 )}_{m}(i\sinh x)}\bigg )^2. \nonumber\\
\ea
The bound state spectrum of this extended potential is the same (isospectral) 
as that of the conventional one but the 
eigenfunctions are different and written in term of exceptional $X_m$ Jacobi 
polynomials $\hat{P}^{(\alpha ,\beta )}_{n +m} (i\sinh x)$ as
\be\label{scarfextdwf}
\psi^{(1)}_{ext,n,m}(x,a,b)\propto \frac{\sech^a x\exp[-ib\tan^{-1}(\sinh x)]}{P^{(-\alpha-1 ,\beta-1 )}_{m} (i\sinh x)}
\hat{P}^{(\alpha ,\beta )}_{n +m} (i\sinh x),
\ee
where 
\ba
\hat{P}^{(\alpha,\beta)}_{n+m}(z(x))=(-1)^m\bigg[\frac{(1+\alpha+\beta+n)}{2(1+\alpha+n)}(z(x)-1)P^{(-\alpha-1,\beta-1)}_{m}(g)P^{(\alpha+2,\beta)}_{n-1}(z(x))\nonumber \\+\frac{(1+\alpha-m)}{(\alpha+1+n)}P^{(-2-\alpha,\beta)}_{m}(z(x))P^{(\alpha+1,\beta-1)}_{n}(z(x))\bigg];\quad n,m\geq 0.
\ea
is the exceptional Jacobi polynomial. 
Similar to the Scarf-II potential, this extended
potential is also SI under the translation of the parameters
$a \rightarrow (a-1)$. 
The transmission and reflection amplitudes for this potential are known 
\cite{scattpt} and are given by
\be\label{tleftxm}
T^{(1)}_{ext,scarf}(k,m,a,b)= T^{(1)}_{scarf}(k,a,b)\zeta(m,a,b)\,,
\ee
and
\be\label{rleftxm}
R^{(1)}_{ext,scarf}(k,m,a,b)= R^{(1)}_{scarf}(k,a,b)\zeta(m,a,b),
\ee
where $\zeta(m,a,b)=\left(\frac{[b^2-(ik-\frac{1}{2})^2]+(b-ik+\frac{1}{2})(1-m)}
{[b^2-(ik+\frac{1}{2})^2]+(b+ik+\frac{1}{2})(1-m)}\right)$. 

Remarkably, it turns out that unlike the conventional Scarf-II 
potential (\ref{spot1}), the corresponding rationally extended Scarf-II 
potential (\ref{scarfextdpot}) is not invariant under the parametric 
transformation $b\longleftrightarrow a+\frac{1}{2}$, rather it generates 
another extended potential \cite{para} given by
\be\label{extdscarfp}
{V}^{(1,p)}_{m,ext}(x,a,b)= V^{(1)}_{m,ext}(x,b\leftrightarrow a+\frac{1}{2}).
\ee
The energy eigenvalues of this potential are isospectral to that of the
conventional potential obtained after the transformation
$b\longleftrightarrow a+\frac{1}{2}$ given by Eq. (\ref{engp1})
and the eigenfunction is 
\be\label{extdscarfwf1}
\psi^{(1,p)}_{ext,n,m}(x,a,b)= \psi^{(1)}_{ext,n,m}(x,a\rightarrow b-1/2,b\rightarrow a+1/2).
\ee
This potential (\ref{extdscarfp}) is also SI under the translation of parameter
$b\longrightarrow b-1$. Since the potentials obtained after parametric 
transformations are different, hence the scattering amplitudes corresponding 
to these potentials are also different which are given by  
\ba
T^{(1,p)}_{ext,scarf}(k,m,a,b)&=& T^{(1)}_{ext,scarf}(k,m,a\rightarrow b-\frac{1}{2},b\rightarrow a+\frac{1}{2})\nonumber\\
R^{(1,p)}_{ext,scarf}(k,m,a,b)&=& R^{(1)}_{ext,scarf}(k,m,a\rightarrow b-\frac{1}{2},b\rightarrow a+\frac{1}{2}).
\ea
Thus, it turns out that 
the extended potentials do not respect the parametric symmetry. As a result for a given value of $[a,b]$ 
(both integers or half integers) one has in fact 
two different sets of rationally extended potentials for a given $m$ which are 
reflectionless. The only exceptions to these are the cases when either 
$[a=(2N-1)/2,b=1/2]$ or $[a=0,b=N]$ where only one rational 
partner exists. Thus in all, one has $2(2N-1)m+2N$ number of complex 
PT-invariant reflectionless potentials, with the $2N$ potentials being 
the nonrational (or $m =0$) ones.

\section{Illustration For Three Bound States ($N=3$)}

In this section, as an illustration, we consider all reflectionless potentials 
(both real and complex PT-invariant ones) with three bound states and show the 
behavior of these potentials and their corresponding normalized ground state 
eigenfunctions through graphical representation.

\subsection{Real reflectionless potential}

In this case, we fix the value of parameter $N=3$ which gives the potential 
(\ref{real_pot})
 \be\label{rpot1}
V^{(1)}(x)=-12\sech^{2}(x),
\ee 
with three bound states with the corresponding binding energies being 
$E^{(1)}_{0}=-9, E^{(1)}_{1}=-4$ and $E^{(1)}_{2}=-1$. The normalized ground state is 
\be\label{ref2} 
\psi^{(1)}_{0} (x)= \frac{ \sqrt{15}}{4} \sech^3 x\,.
\ee
Thus the partner potential with its normalized ground state
eigenfunction are
\be
V^{(2)}(x) = -6\sech^2(x)\,,~~\psi_{0}^{(2)}(x) = \frac{\sqrt{3}}{2} \sech^2(x)\,.
\ee
It is straight forward to calculate the integral (\ref{intim}) using the 
potential (\ref{rpot1}) and one obtains
\be
I(x)=\frac{1}{16} \bigg(8 + [8 + 4 \sech^2(x) + 3\sech^4(x)] \tanh(x) \bigg)\,,
\ee
which gives one continuous parameter family of real reflectionless potentials 
\ba\label{opA3}
\hat{V}^{(1)}(x,\lambda) &=&6 \sech^2(x)\bigg[-2+\frac{1}{(8 + 16 \lambda + (8 + 4 \sech^2(x) + 3 \sech^4(x)) \tanh(x))^2}\nonumber\\
&\times &[15 \sech^4(x) \{ (1 + 3 \cosh(2x) + \cosh(4x) ) \sech^6(x) \nonumber\\
&+& 16 \tanh(x) (1 + 2 \lambda + \tanh(x) )\}]\bigg].
\ea
The normalized ground state eigenfunction for this potential is obtained as
\ba\label{gswfop2}
\hat{\psi}^{(1)}_{0}(x,\lambda)=\frac{4[15\lambda (\lambda+1)]^{\frac{1}{2}} \sech^3(x)}{(8 + 
 16 \lambda + (8 + 4 \sech^2(x) + 3 \sech^4(x) ) \tanh(x) )}.
\ea 
In the limit of $\lambda \rightarrow 0$ and $-1$, we get the reflectionless 
Pursey and the AM potentials respectively with two bound states. The 
expressions for these two potentials are given by 
\be\label{pa3}
V^{[P/
AM]}(x)=-\frac{24 \sech^2(x) [25 \cosh(2 x) + 13 \cosh(4 x) \mp 3 (\mp 5 + 5 \sinh(2 x) + 4 \sinh(4 x))]}{(5 + 11 \cosh (2 x) \mp 9 \sinh^2 (2 x))}, 
\ee 
where upper sign corresponds to the Pursey and the lower one for the 
AM potential. The plots of $\hat{V}^{(1)}(x,\lambda)$ for positive and negative 
$\lambda$ are shown in Fig. 1(a) and 1(b) respectively. The AM ($V^{[AM]}(x)$), 
the Pursey ($V^{[P]}(x)$) and the partner potential ($V^{(2)}(x)$) are shown in 
Fig. 1(c). The normalized ground state wavefunctions for some positive 
values of $\lambda$ are also shown in Fig. 1(d).\\
\includegraphics[scale=1.0]{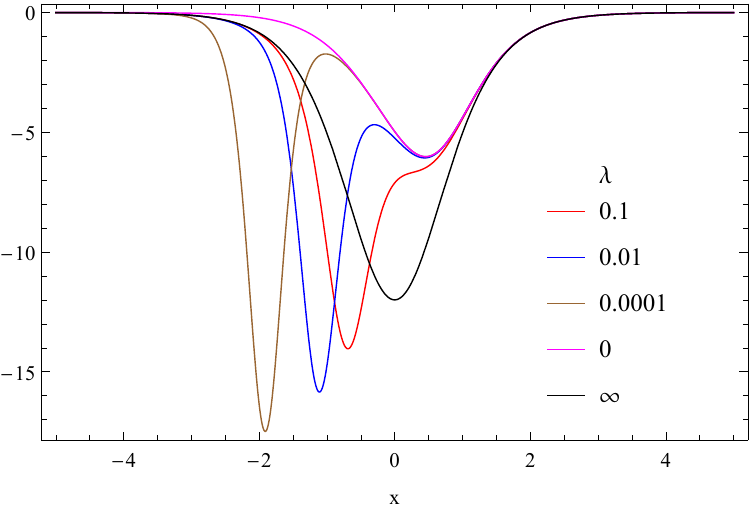}\\
{\bf Fig.1}{(a) {\it One-parameter family of potential $\hat{V}^{(1)}
(x,\lambda)$  for positive $\lambda=0.1,0.01,0.0001,0$ and $\infty$. The Pursey
potential is shown for $\lambda=0$.}\\
\includegraphics[scale=1.0]{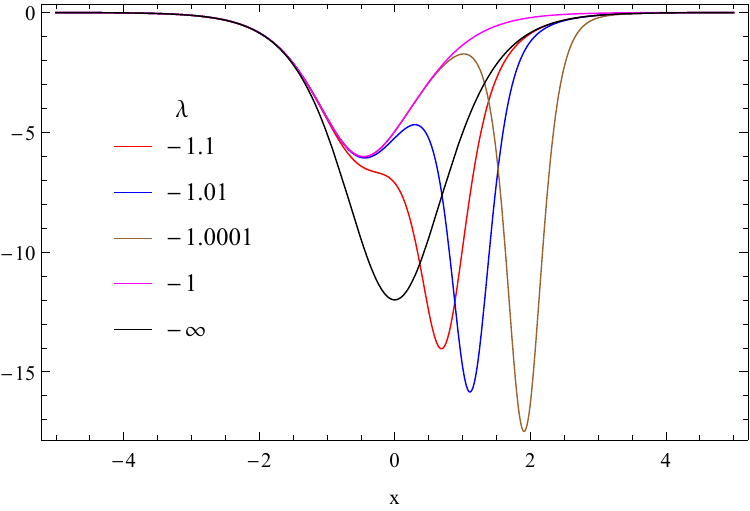}\\
{\bf Fig.1}{(b) {\it One-parameter family of potential $\hat{V}^{(1)}
(x,\lambda)$  for negative $\lambda=-1.1,-1.01,-1.0001,-1$ and $-\infty$. The 
AM potential is shown for $\lambda=-1$.}\\\ 
 \includegraphics[scale=1.0]{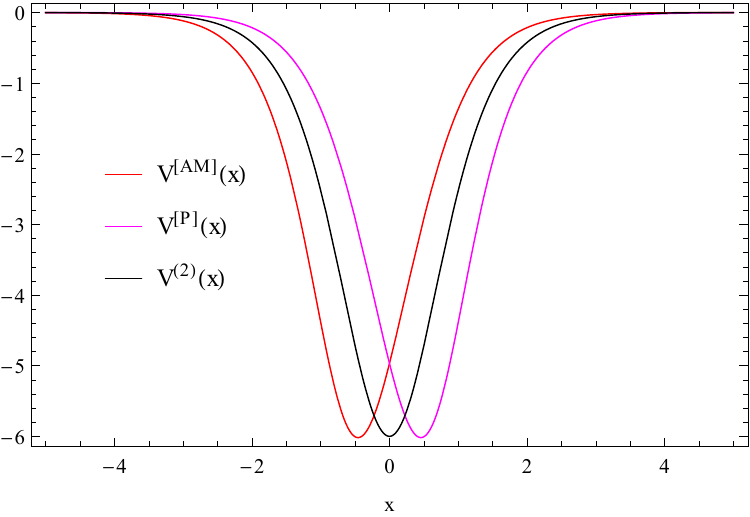}\\
{\bf Fig.1}{(c) {\it The Pursey potential ${V}^{[P]}(x)$, the AM potential 
${V}^{[AM]}(x)$ 
and the partner potential ${V}^{(2)}(x)$.}\\
\includegraphics[scale=1.0]{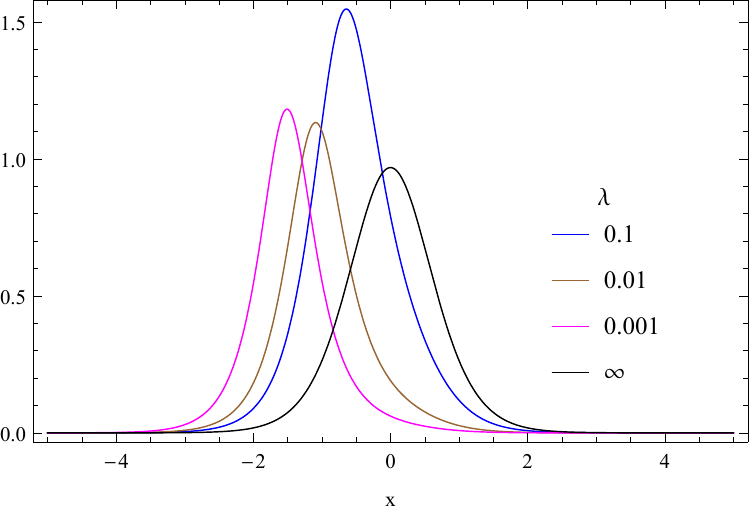}\\
{\bf Fig.1}{(d) {\it Normalized ground-state wavefunctions 
$\hat{\psi}^{(1)}_{0}(x,\lambda )$ for some potentials 
 (with positive $\lambda=0.1,0.01,0.001$ and $\infty$.)}\\

 
 

\subsection{PT symmetric complex Scarf-II potentials} 

\subsubsection{ The conventional potential}

In this case, for three bound states ($N=3$), we have six different possible 
combinations of $[a,b]$ given by 
\be
[a,b]=[5/2,1/2], [3/2,3/2], [1/2,5/2], [2,1], [1,2], [0,3]\,,
\ee
and hence we have six reflectionless potentials. If we plot the potentials for 
these combinations of $a$ and $b$, while the potentials $[5/2,1/2]$, $[3/2,3/2]$ and $[1/2,5/2]$ are same as that of 
$[0,3]$, $[1,2]$ and $[2,1]$ respectively as can be seen from the plots, the corresponding
 eigenfunctions are different for half-integer and integer combinations of 
both $a$ and $b$. As mentioned in Eqs. (\ref{hint}) and (\ref{int}), the 
integer combination $[0,3]$ is not acceptable for the potential $V^{(1)}(x,a,b)$, however this is acceptable for 
the parametric case $V^{(1,p)}(x,a,b)$. Similarly, the first combination $[5/2,1/2]$  is well acceptable for the first potential, but 
not for the parametric case.
The plots of these potentials (real and imaginary parts) 
are shown in Fig. $2$. The corresponding eigenfunctions with their parametric forms are also 
shown in Fig. $3$ and $4$ receptively. We also compare the eigenfunctions of conventional $PT$ symmetric potentials with their parametric 
counterparts graphically (shown in Fig. $5$).   \\
\begin{figure}[ht]
{\begin{minipage}[c][1\width]{0.3\textwidth}
\centering
 \includegraphics[scale=.38]{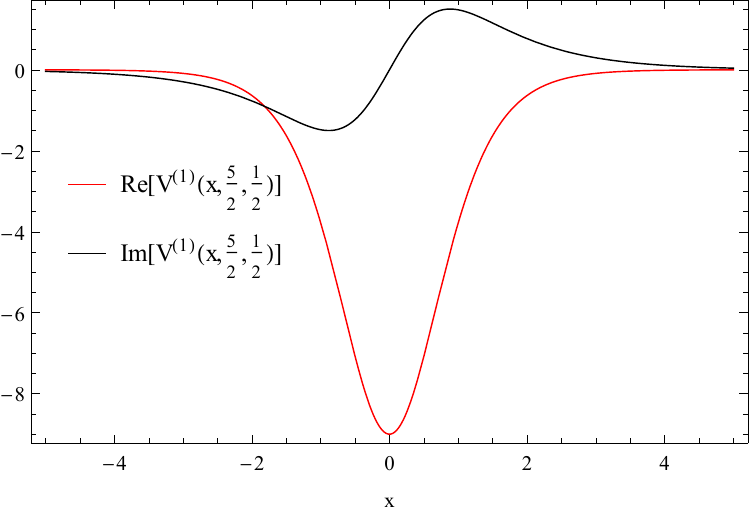}
 \subcaption{$a=\frac{5}{2},b=\frac{1}{2}$}
\end{minipage}}
\hfill
 {\begin{minipage}[c][1\width]{0.3\textwidth}
\centering
 \includegraphics[scale=.38]{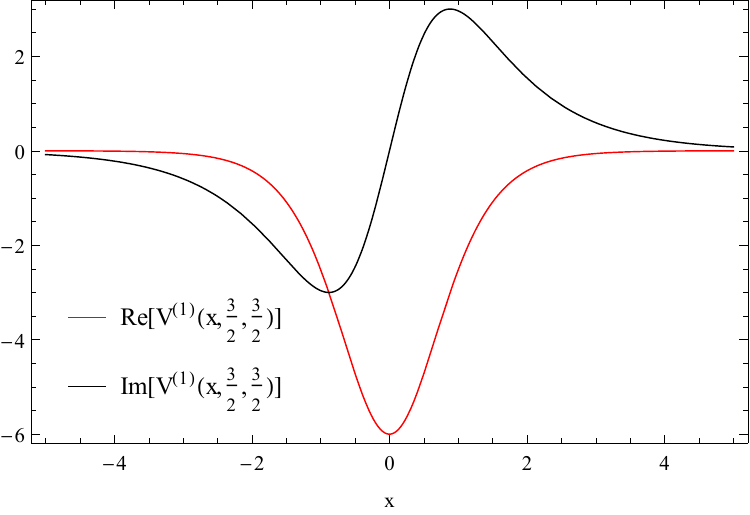}
 \subcaption{$a=\frac{3}{2}, b=\frac{3}{2}$}
\end{minipage}}
\hfill
{\begin{minipage}[c][1\width]{0.3\textwidth}
\centering
\includegraphics[scale=.38]{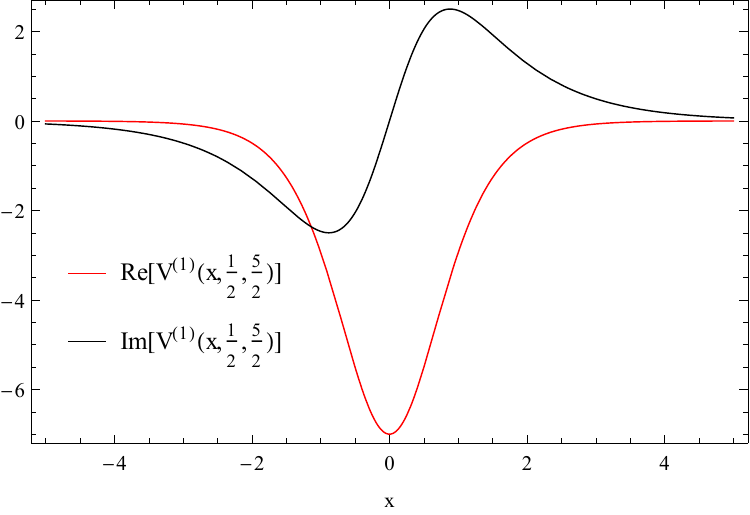}
\subcaption{$a=\frac{1}{2}, b=\frac{5}{2}$}
\end{minipage}}\\
{\bf Fig.2}: {(a)-(c) Real and imaginary parts of the conventional $PT$ symmetric Scarf-II potential ($V^{(1)}(x,a,b)$) vs $x$ for different $a$ and $b$.}
\end{figure}
\begin{figure}[ht]
{\begin{minipage}[c][1\width]{0.3\textwidth}
\centering
 \includegraphics[scale=.38]{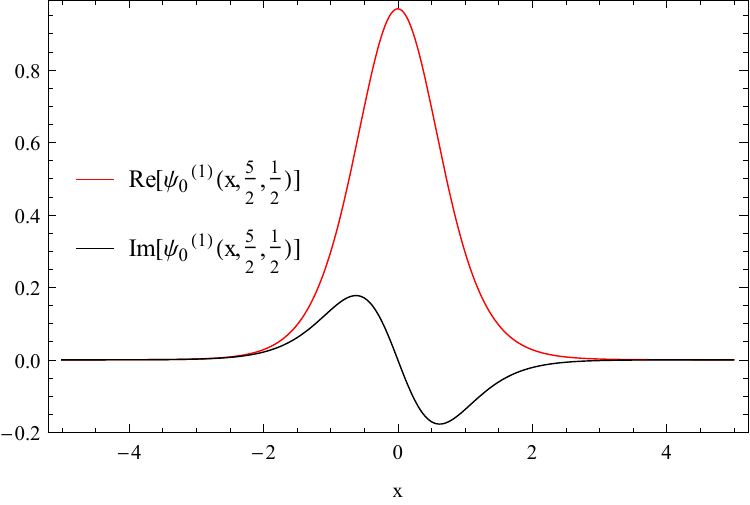}
 (a) {$a=\frac{5}{2},b=\frac{1}{2}$}
\end{minipage}}
\hfill
 {\begin{minipage}[c][1\width]{0.3\textwidth}
\centering
 \includegraphics[scale=.38]{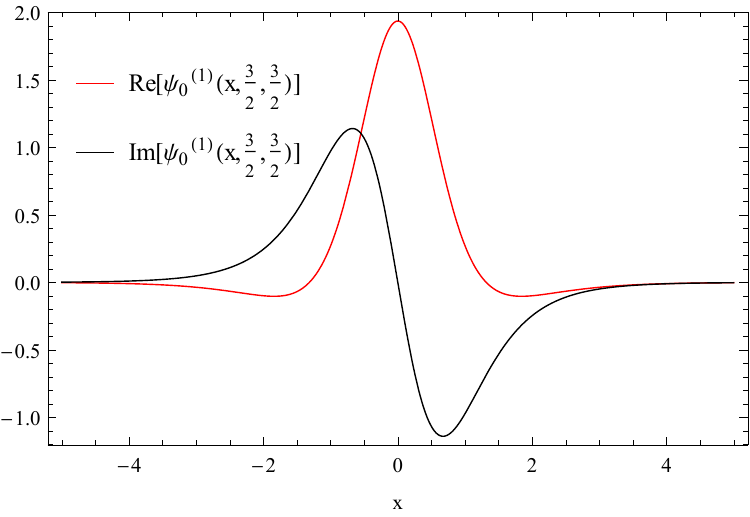}
 (b) {$a=\frac{3}{2}, b=\frac{3}{2}$}
\end{minipage}}
\hfill
{\begin{minipage}[c][1\width]{0.3\textwidth}
\centering
\includegraphics[scale=.38]{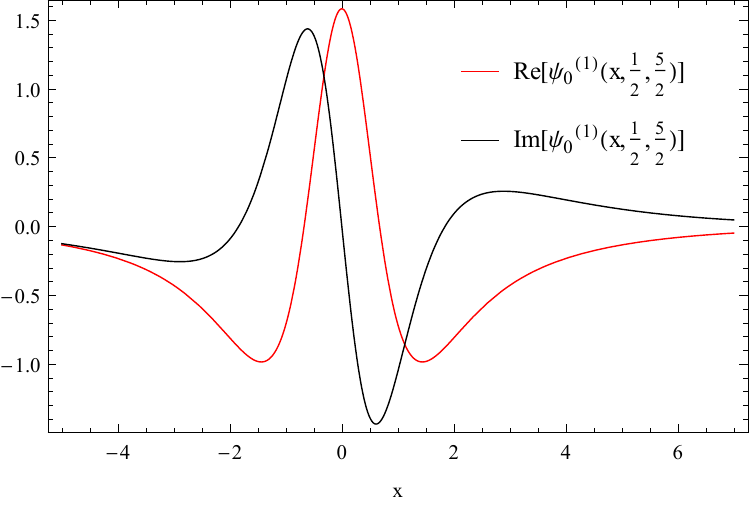}
(c) {$a=\frac{1}{2}, b=\frac{5}{2}$}
\end{minipage}}\\
{\bf Fig.3 (i)}: {(a)-(c) Real and imaginary parts of the normalized ground state eigenfunctions $\psi^{(1)}_0(x,a,b)$ vs $x$ for half-integer values
of $a$ and $b$.}
\end{figure}
\begin{figure}[ht]
{\begin{minipage}[c][1\width]{0.45\textwidth}
\centering
 \includegraphics[scale=.52]{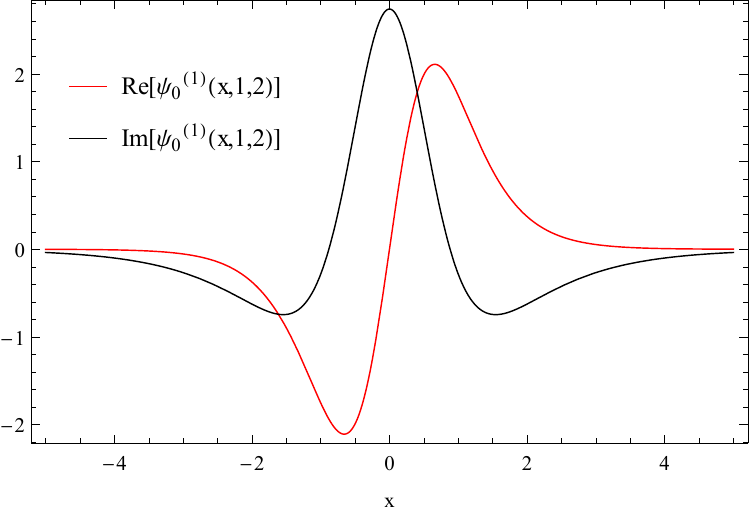}
 (a) {$a=1,b=2$}
\end{minipage}}
\hfill
 {\begin{minipage}[c][1\width]{0.45\textwidth}
\centering
 \includegraphics[scale=.52]{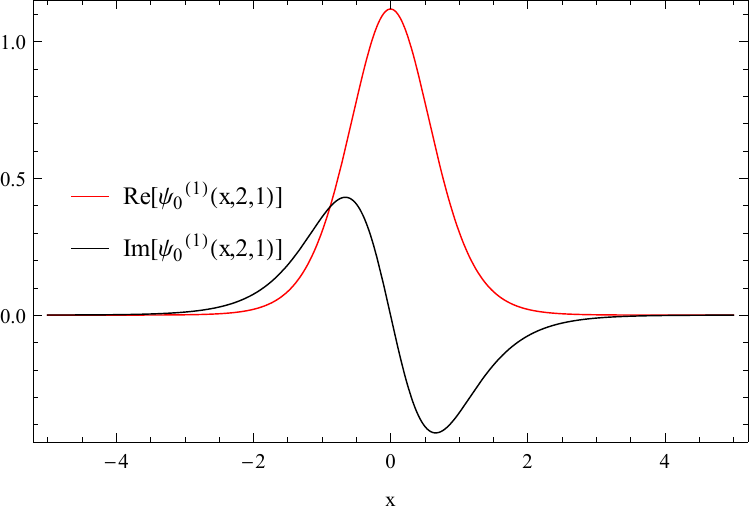}
 (b) {$a=2, b=1$}
\end{minipage}}\\
{\bf Fig.3 (ii)}: {(a)-(b) Real and imaginary parts of the normalized ground state eigenfunctions for $\psi^{(1)}_0(x,a,b)$ vs $x$ for integer values
of $a$ and $b$.}
\end{figure}
\begin{figure}[ht]
{\begin{minipage}[c][1\width]{0.45\textwidth}
\centering
 \includegraphics[scale=.52]{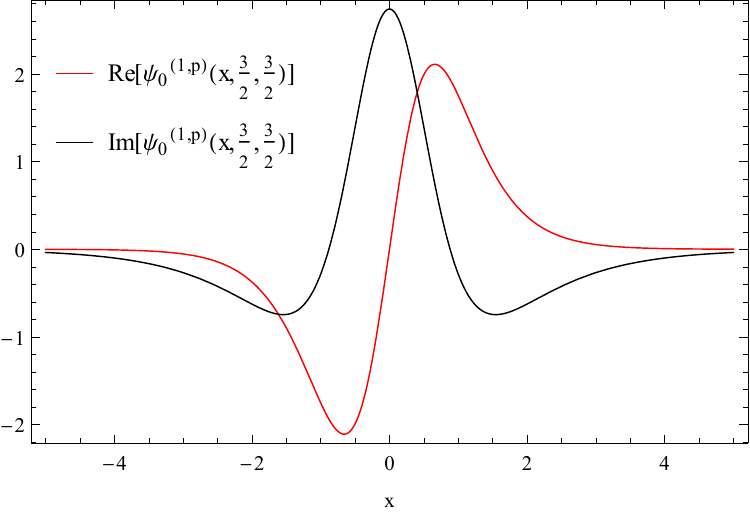}
 (a) {$a=\frac{3}{2},b=\frac{3}{2}$}
\end{minipage}}
\hfill
 {\begin{minipage}[c][1\width]{0.45\textwidth}
\centering
 \includegraphics[scale=.52]{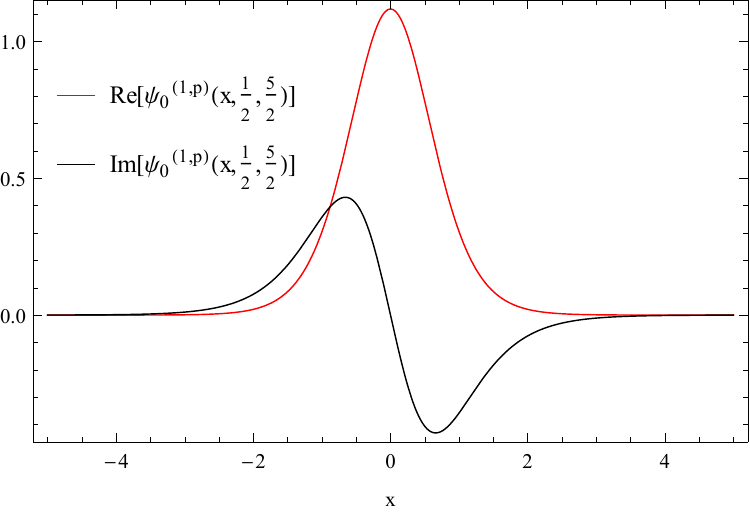}
 (b) {$a=\frac{1}{2}, b=\frac{5}{2}$}
\end{minipage}}\\
{\bf Fig.4 (i)}: {(a)-(b) Real and imaginary parts of the normalized ground state eigenfunctions (for half-integer values of $a$ and $b$) for the parametric case $\psi^{(1,p)}_0(x,a,b)$.}
\end{figure}
\begin{figure}[ht]
{\begin{minipage}[c][1\width]{0.3\textwidth}
\centering
 \includegraphics[scale=.38]{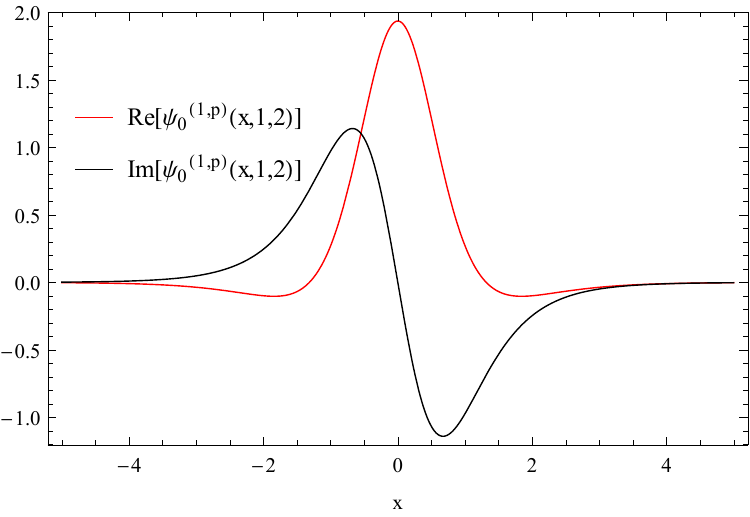}
 (a) {$a=1,b=2$}
\end{minipage}}
\hfill
 {\begin{minipage}[c][1\width]{0.3\textwidth}
\centering
 \includegraphics[scale=.38]{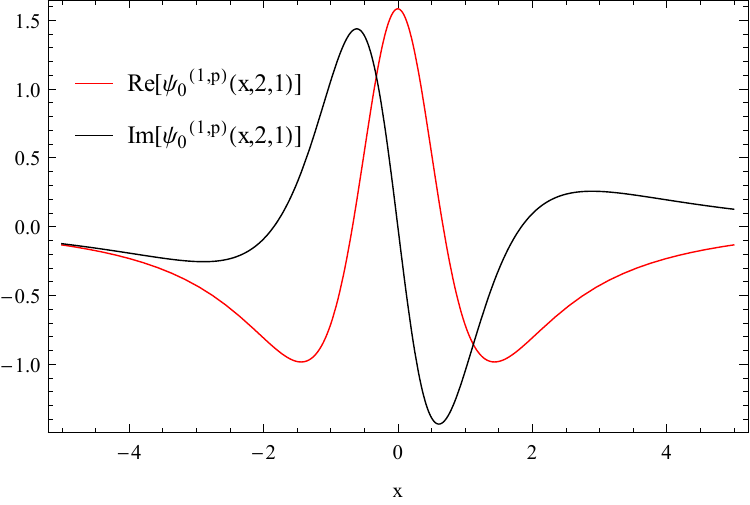}
 (b) {$a=2, b=1$}
\end{minipage}}
\hfill
 {\begin{minipage}[c][1\width]{0.3\textwidth}
\centering
 \includegraphics[scale=.38]{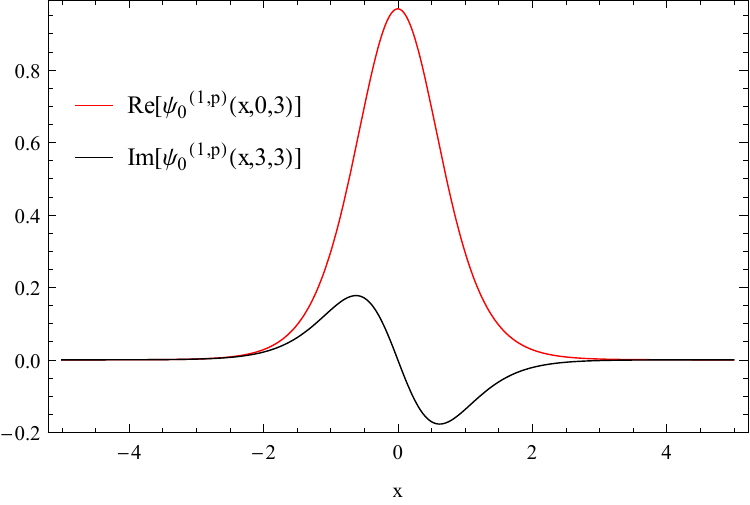}
 (b) {$a=0, b=3$}
\end{minipage}}\\
{\bf Fig.4 (ii)}: {(a)-(b) Real and imaginary parts of the normalized ground state eigenfunctions (for integer values of $a$ and $b$) for the parametric case $\psi^{(1,p)}_0(x,a,b)$.}
\end{figure}
\begin{figure}[ht]
{\begin{minipage}[c][1\width]{0.45\textwidth}
\centering
 \includegraphics[scale=.52]{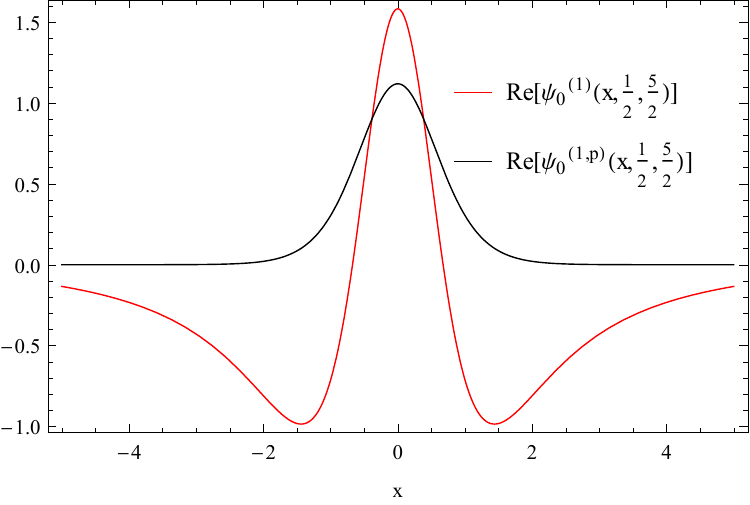}
 (a) 
\end{minipage}}
\hfill
 {\begin{minipage}[c][1\width]{0.45\textwidth}
\centering
 \includegraphics[scale=.52]{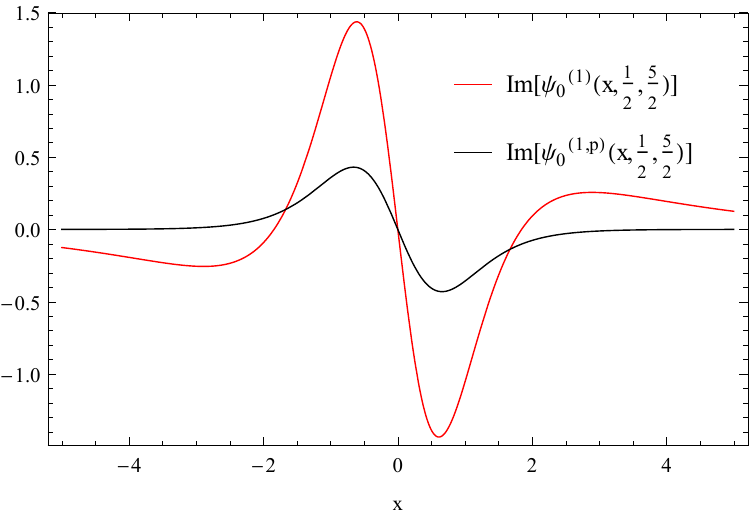}
 (b) 
\end{minipage}}\\
{\bf Fig.5 (i)}: {(a)-(b) Comparison of real and imaginary parts of the normalized ground state eigenfunctions for the conventional $\psi^{(1)}_0(x,a,b)$ and parametric cases $\psi^{(1,p)}_0(x,a,b)$ for $a=\frac{1}{2}, b=\frac{5}{2}$.}
\end{figure}
\begin{figure}[ht]
{\begin{minipage}[c][1\width]{0.45\textwidth}
\centering
 \includegraphics[scale=.52]{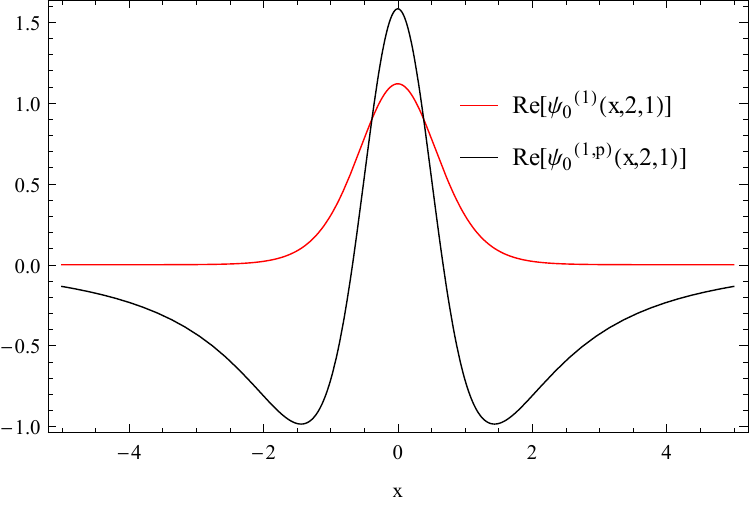}
 (a) 
\end{minipage}}
\hfill
 {\begin{minipage}[c][1\width]{0.45\textwidth}
\centering
 \includegraphics[scale=.52]{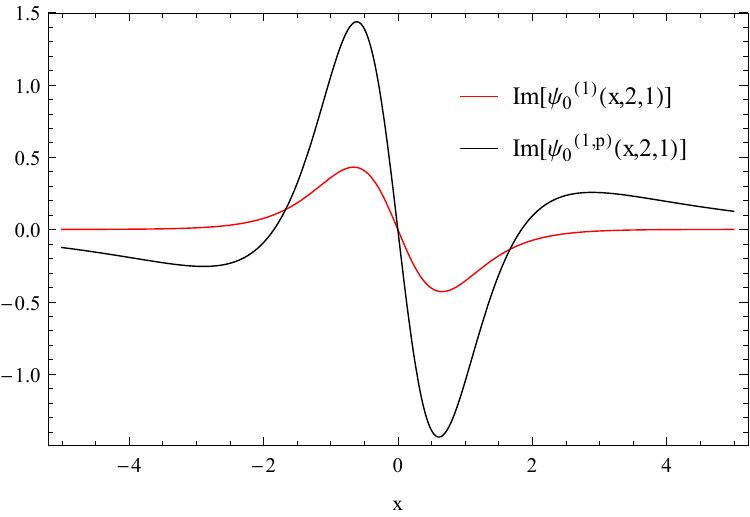}
 (b) 
\end{minipage}}\\
{\bf Fig.5 (ii)}: {(a)-(b) Comparison of real and imaginary parts of the normalized ground state eigenfunctions for the conventional $\psi^{(1)}_0(x,a,b)$ and parametric cases $\psi^{(1,p)}_0(x,a,b)$ for $a=2, b=1$.}
\end{figure}

\FloatBarrier
\subsubsection{Rationally extended $PT$ symmetric Scarf-II potential}

In this case, unlike the conventional $PT$ symmetric Scarf-II potential, the extended potentials as well as the corresponding 
eigenfunctions are completely different under parametric transformations. For $m=1$, the expression of potential and the normalizable eigenfunction 
are given as 
\be
V^{(1)}_{1,ext}(x,a,b)=V^{(1)}(x,a,b)+\frac{(-2(2a+1))}{(-2ib\sinh(x) + 
    2a+1)} + \frac{(2((2a+1)^2-4b^2))}{(-2ib\sinh(x)+2a+1)^2}
\ee
and 
\be\label{scarfwfx1}
\psi^{(1)}_{ext,n,m}(x,a,b)\propto \frac{\sech^a x\exp[-ib\tan^{-1}(\sinh x)]}{(-2ib\sinh(x)+2a+1)}
\hat{P}^{(\alpha ,\beta )}_{n+1} (i\sinh x),
\ee
where $\hat{P}^{(\alpha ,\beta )}_{n+1} (i\sinh x)$ is the $X_1$ exceptional Jacobi polynomial.

We consider the same sets of parameters $[a,b]$ (half-integers as well as integers) as discussed 
in the above conventional case and show the behaviors of potentials 
$V^{(1)}_{1,ext}(x,a,b)$, $V^{(1,p)}_{1,ext}(x,a,b)$ and the corresponding 
eigenfunctions $\psi^{(1)}_{1,ext}(x,a,b)$, $\psi^{(1,p)}_{1,ext}(x,a,b)$ 
respectively in Figs. $6$, $7$ and $8$. 
\begin{figure}[ht]
{\begin{minipage}[c][1\width]{0.3\textwidth}
\centering
 \includegraphics[scale=.38]{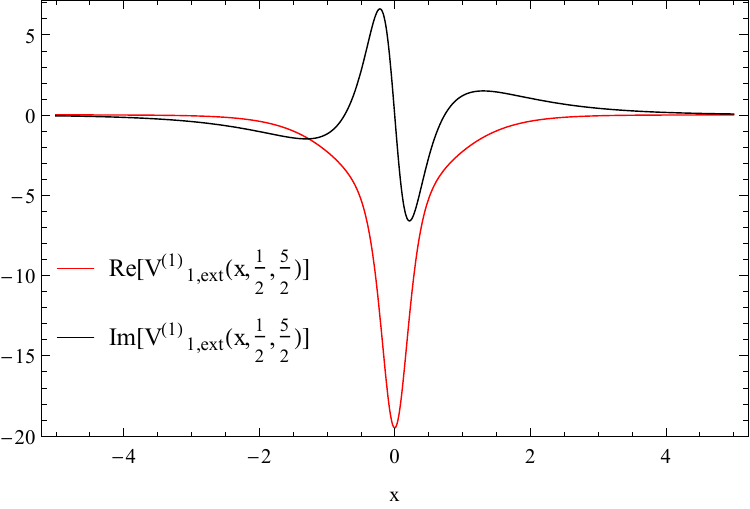}
\end{minipage}}
\hfill
 {\begin{minipage}[c][1\width]{0.3\textwidth}
\centering
 \includegraphics[scale=.38]{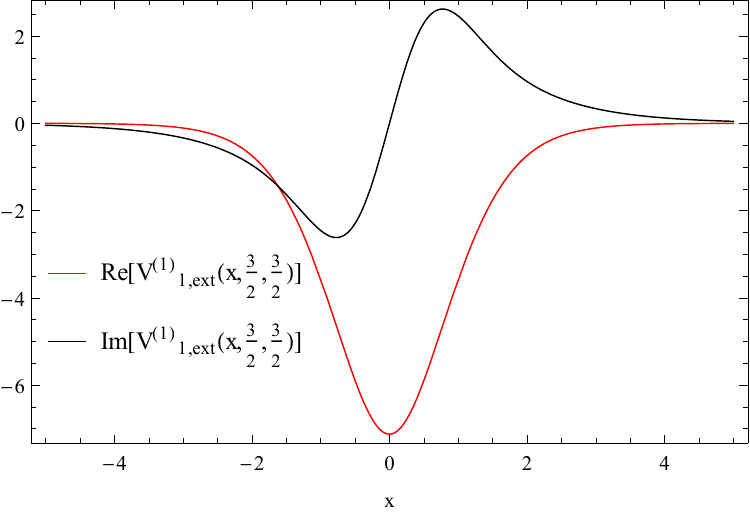}
\end{minipage}}
\hfill
{\begin{minipage}[c][1\width]{0.3\textwidth}
\centering
\includegraphics[scale=.38]{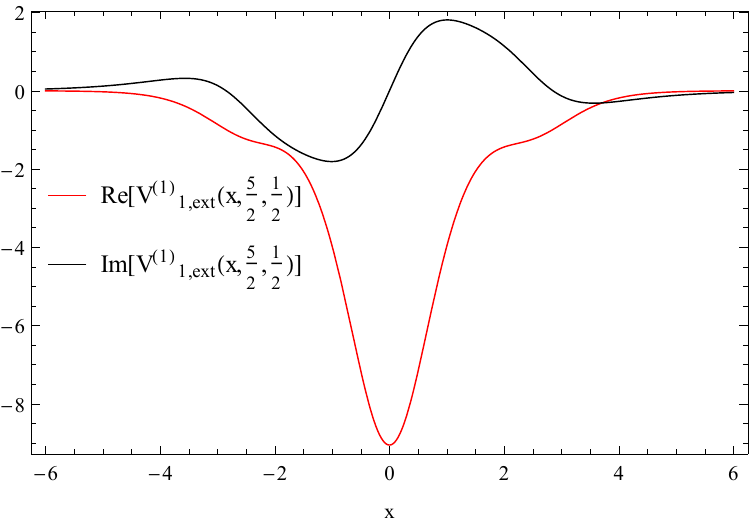}
\end{minipage}}
\hfill
{\begin{minipage}[c][1\width]{0.3\textwidth}
\centering
\includegraphics[scale=.38]{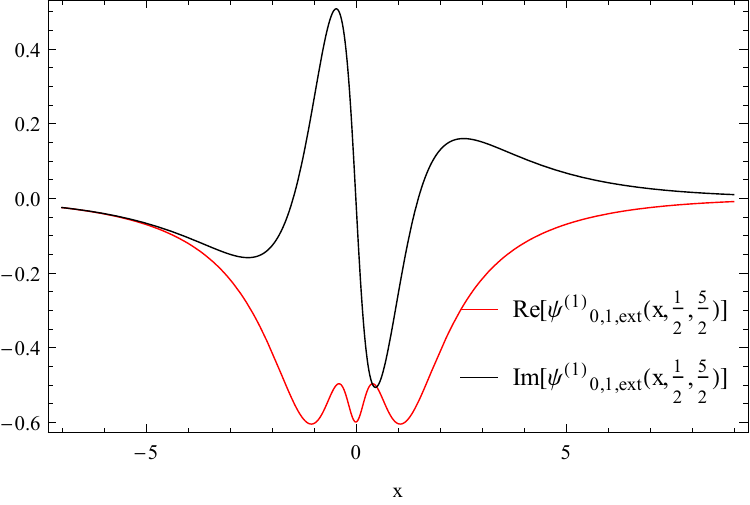}
(a) {$a=\frac{1}{2}, b=\frac{5}{2}$}
\end{minipage}}
\hfill
{\begin{minipage}[c][1\width]{0.3\textwidth}
\centering
\includegraphics[scale=.38]{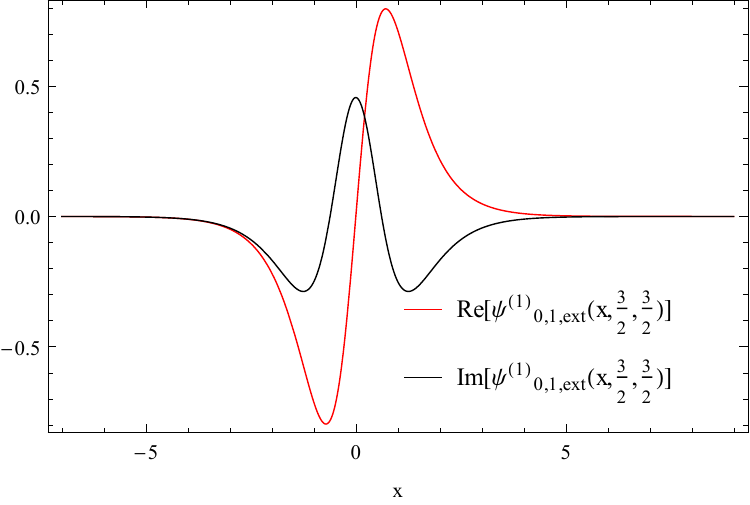}
(b) {$a=\frac{3}{2}, b=\frac{3}{2}$}
\end{minipage}}
\hfill
{\begin{minipage}[c][1\width]{0.3\textwidth}
\centering
\includegraphics[scale=.35]{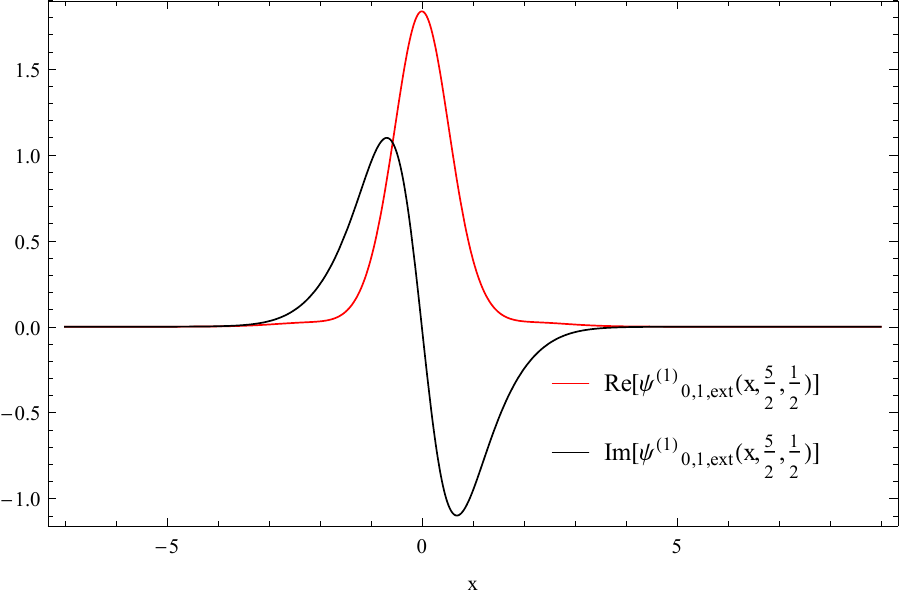}
(c) {$a=\frac{5}{2}, b=\frac{1}{2}$}
\end{minipage}}\\
{\bf Fig.6 (i)}: {(a)-(c) Real and imaginary parts of the RE $PT$ symmetric complex Scarf-II potentials and corresponding eigenfunctions for half-integer values of $a$ and $b$.}\\
\end{figure}
\begin{figure}[ht]
{\begin{minipage}[c][1\width]{0.45\textwidth}
\centering
 \includegraphics[scale=.52]{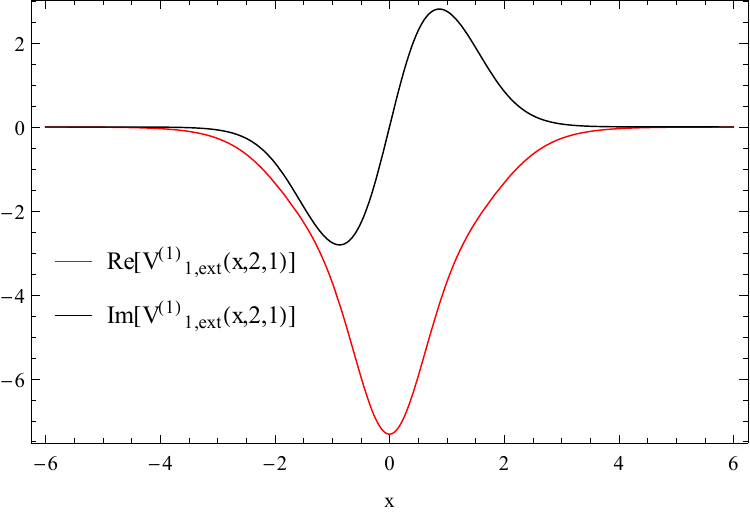}
\end{minipage}}
\hfill
 {\begin{minipage}[c][1\width]{0.45\textwidth}
\centering
 \includegraphics[scale=.52]{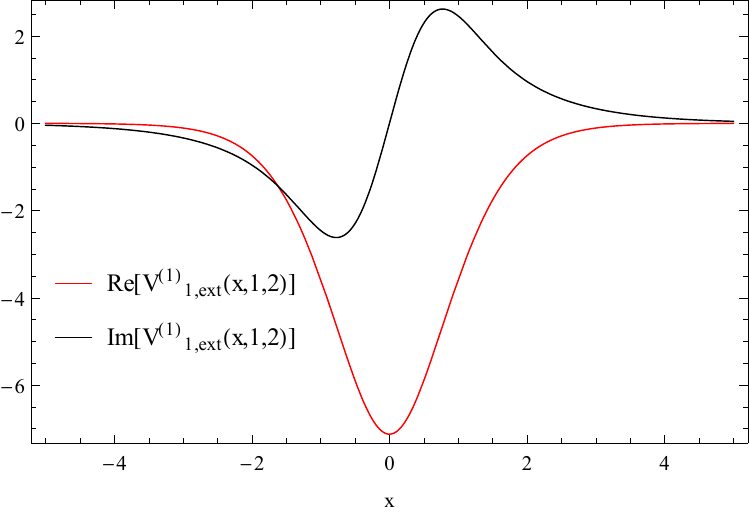}
\end{minipage}}
\hfill
{\begin{minipage}[c][1\width]{0.45\textwidth}
\centering
\includegraphics[scale=.52]{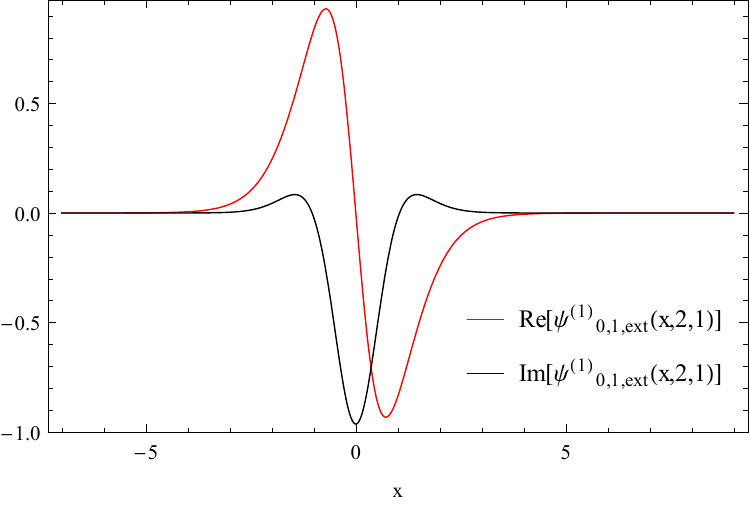}
(a) {$a=2, b=1$}
\end{minipage}}
\hfill
{\begin{minipage}[c][1\width]{0.45\textwidth}
\centering
\includegraphics[scale=.52]{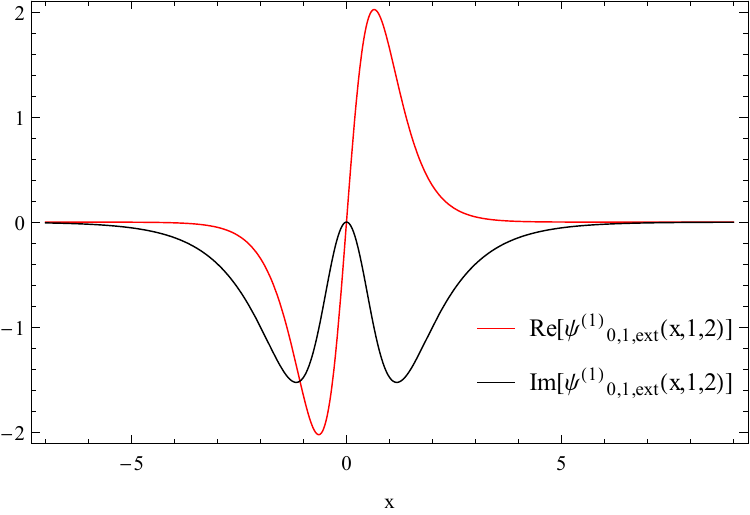}
(b) {$a=1, b=2$}
\end{minipage}}\\
{\bf Fig.6 (ii)}: {(a)-(c) Real and imaginary parts of the RE $PT$ symmetric complex Scarf-II potentials and their corresponding eigenfunctions for integer values of  $a$ and $b$.}\\
\end{figure}
\begin{figure}[ht]
{\begin{minipage}[c][1\width]{0.45\textwidth}
\centering
 \includegraphics[scale=.52]{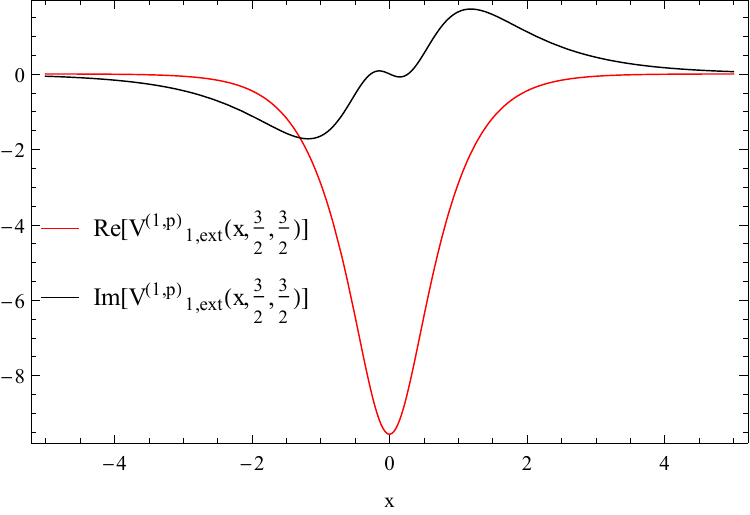}
\end{minipage}}
\hfill
 {\begin{minipage}[c][1\width]{0.45\textwidth}
\centering
 \includegraphics[scale=.52]{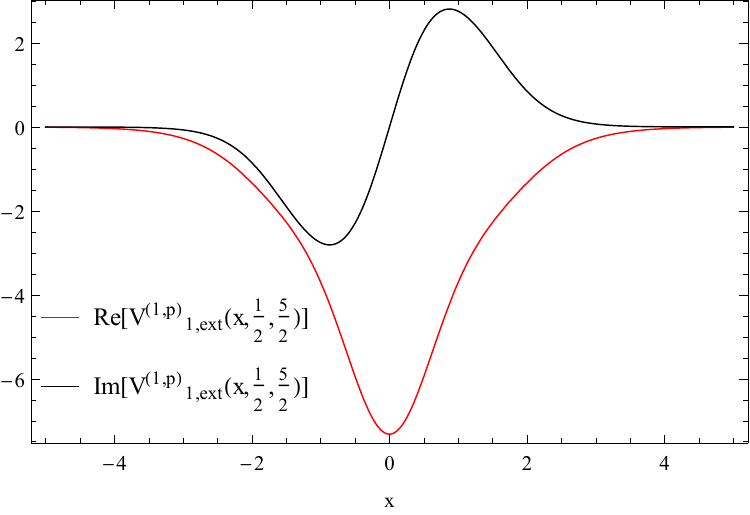}
\end{minipage}}
\hfill
 {\begin{minipage}[c][1\width]{0.45\textwidth}
\centering
 \includegraphics[scale=.52]{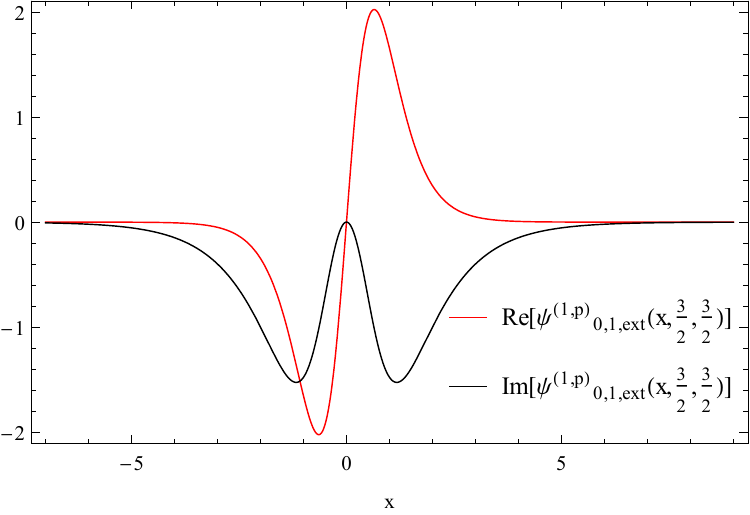}
 (a) {$a=\frac{3}{2}, b=\frac{3}{2}$}
\end{minipage}}
\hfill
 {\begin{minipage}[c][1\width]{0.45\textwidth}
\centering
 \includegraphics[scale=.52]{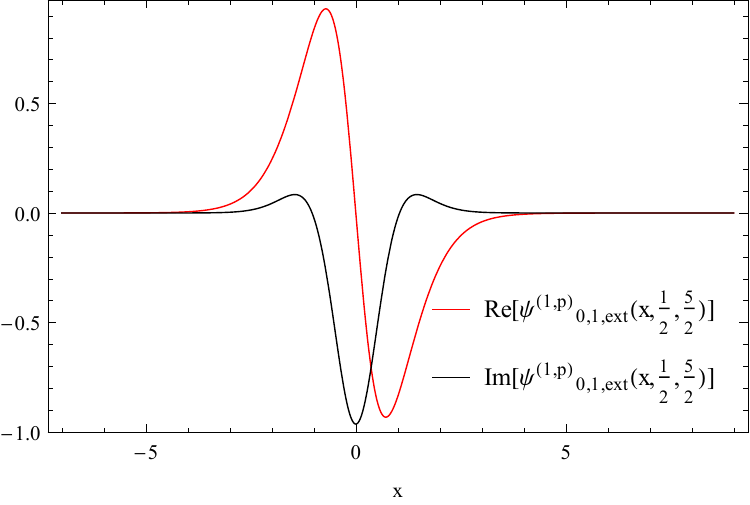}
 (b) {$a=\frac{1}{2}, b=\frac{5}{2}$}
\end{minipage}}\\
{\bf Fig.7 (i)}: {(a)-(b) Real and imaginary parts of the RE $PT$ symmetric complex Scarf-II potentials and their corresponding eigenfunctions obtained after parametric transformation  for half-integer values of $a$ and $b$.}
\end{figure}
\begin{figure}[ht]
{\begin{minipage}[c][1\width]{0.3\textwidth}
\centering
 \includegraphics[scale=.38]{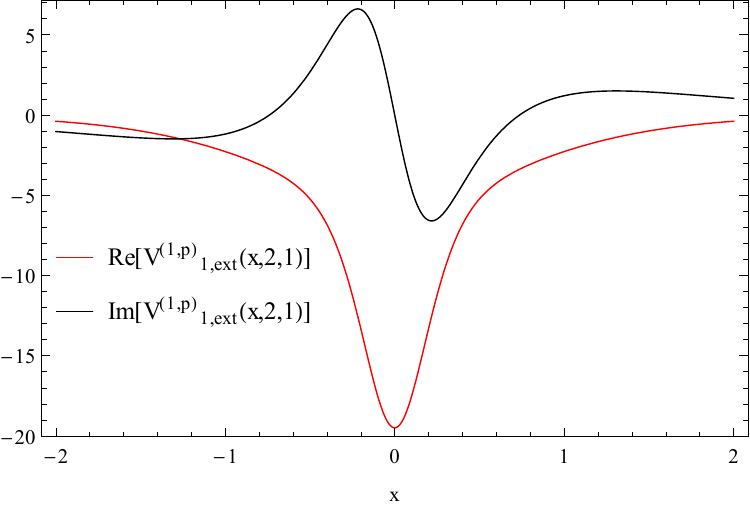}
\end{minipage}}
\hfill
{\begin{minipage}[c][1\width]{0.3\textwidth}
\centering
 \includegraphics[scale=.38]{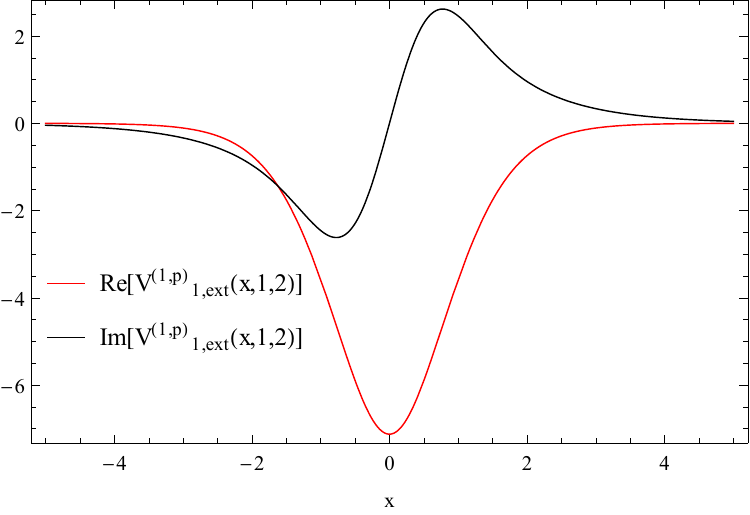}
\end{minipage}}
\hfill
 {\begin{minipage}[c][1\width]{0.3\textwidth}
\centering
 \includegraphics[scale=.38]{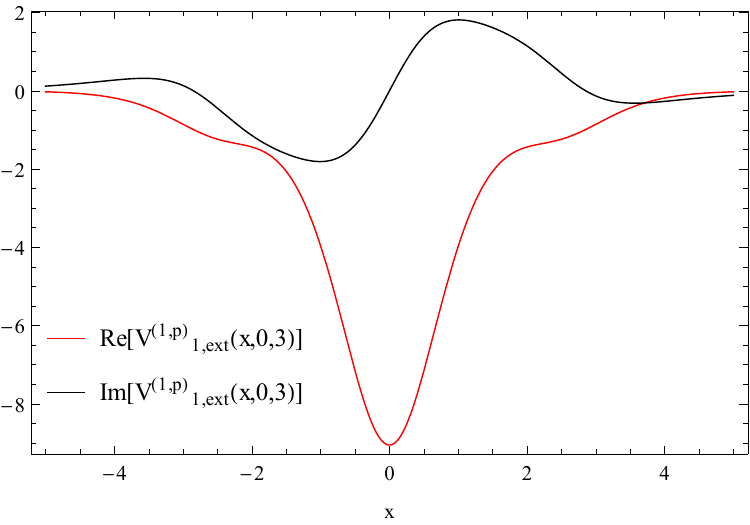}
\end{minipage}}
\hfill
 {\begin{minipage}[c][1\width]{0.3\textwidth}
\centering
 \includegraphics[scale=.38]{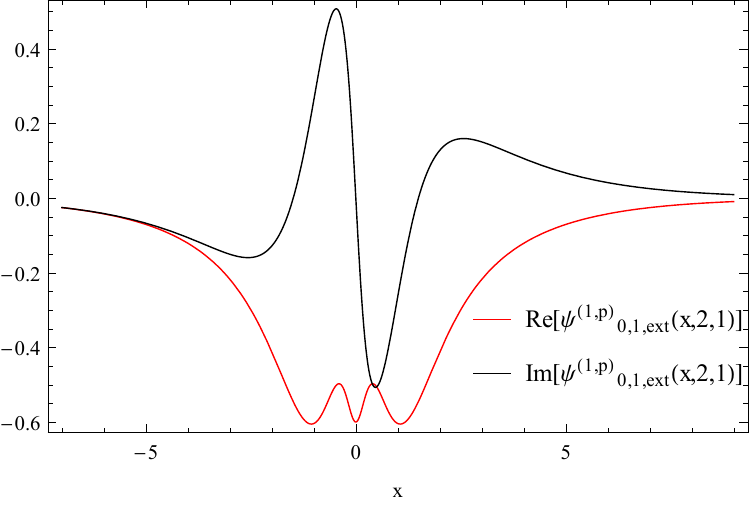}
 (a) {$a=2, b=1$}
\end{minipage}}
\hfill
 {\begin{minipage}[c][1\width]{0.3\textwidth}
\centering
 \includegraphics[scale=.38]{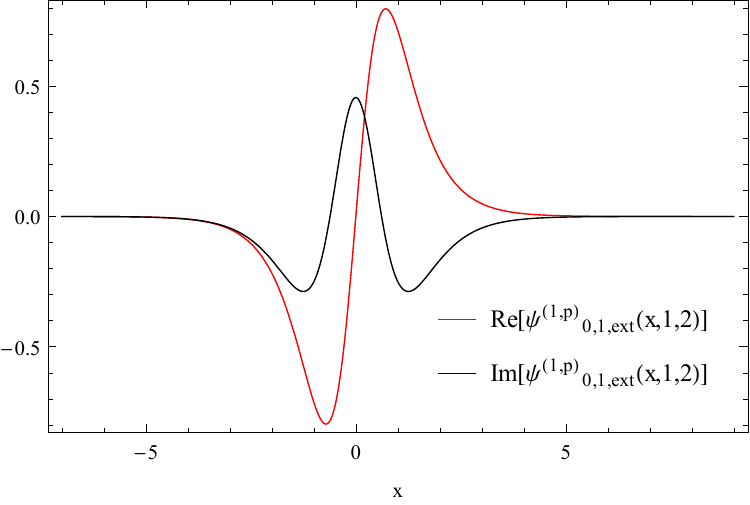}
 (b) {$a=1, b=2$}
\end{minipage}}
\hfill
 {\begin{minipage}[c][1\width]{0.3\textwidth}
\centering
 \includegraphics[scale=.38]{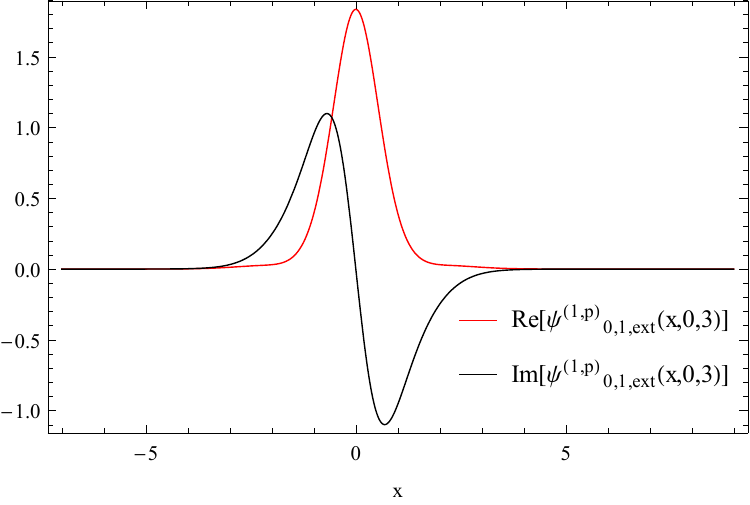}
 (c) {$a=0, b=3$}
\end{minipage}}\\
{\bf Fig.7 (ii)}: {(a)-(b) Real and imaginary parts of the RE $PT$ symmetric complex Scarf-II potentials and their corresponding eigenfunctions obtained after parametric transformation  for integer values of $a$ and $b$.}
\end{figure}

\begin{figure}[ht]
{\begin{minipage}[c][1\width]{0.45\textwidth}
\centering
 \includegraphics[scale=.52]{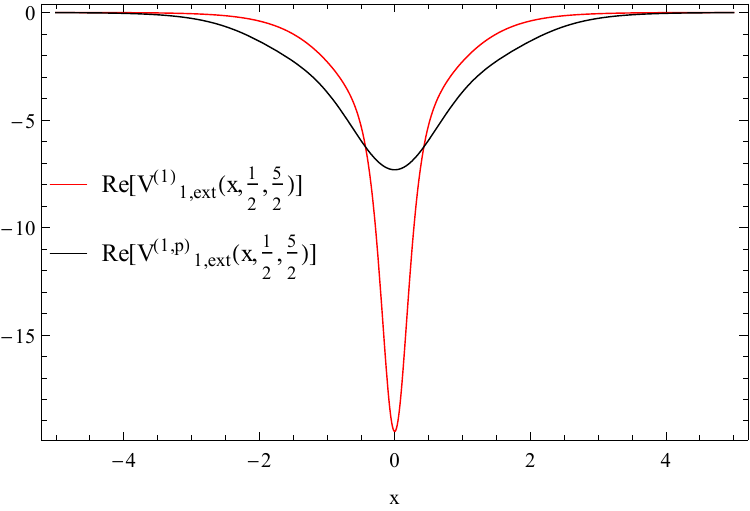}
\end{minipage}}
\hfill
 {\begin{minipage}[c][1\width]{0.45\textwidth}
\centering
 \includegraphics[scale=.52]{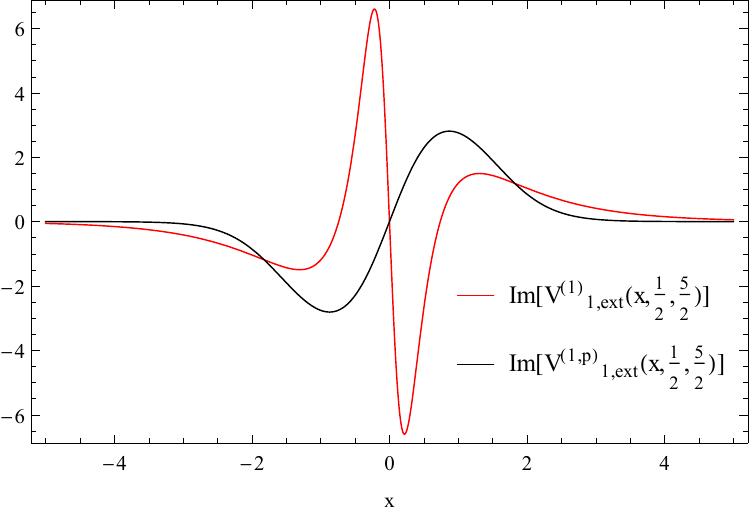}
\end{minipage}}
\hfill
 {\begin{minipage}[c][1\width]{0.45\textwidth}
\centering
 \includegraphics[scale=.52]{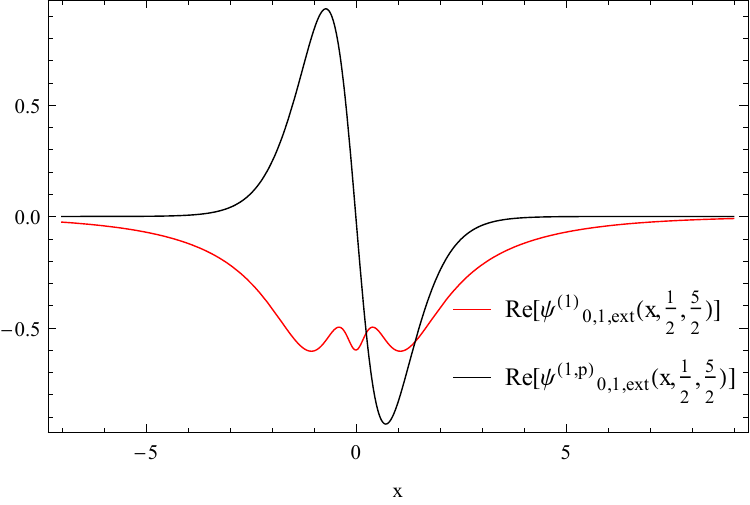}
 (a) $a=\frac{1}{2}, b=\frac{5}{2} $
\end{minipage}}
\hfill
 {\begin{minipage}[c][1\width]{0.45\textwidth}
\centering
 \includegraphics[scale=.52]{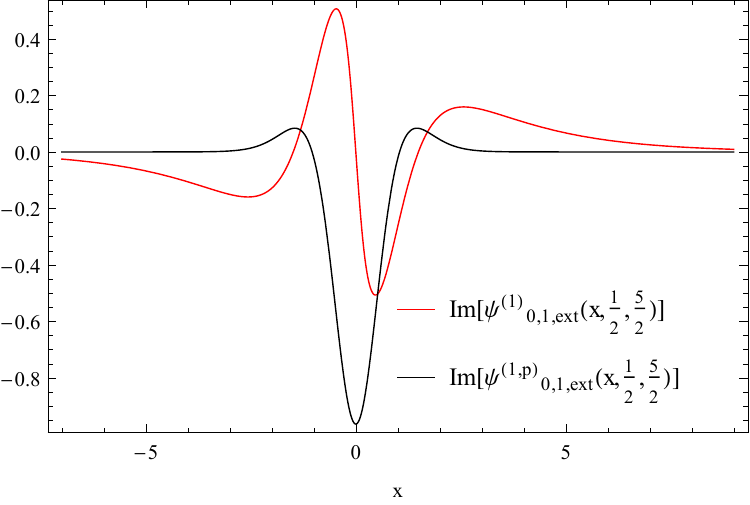}
 (b) $a=\frac{1}{2}, b=\frac{5}{2} $
\end{minipage}}\\
{\bf Fig.8 (i)}: {(a)-(b) Comparison between real and imaginary parts of RE $PT$ symmetric complex Scarf-II potential and their corresponding eigenfunctions obtained after parametric transformation  for half integer combination of $a$ and $b$.}
\end{figure}
\begin{figure}[ht]
{\begin{minipage}[c][1\width]{0.45\textwidth}
\centering
 \includegraphics[scale=.52]{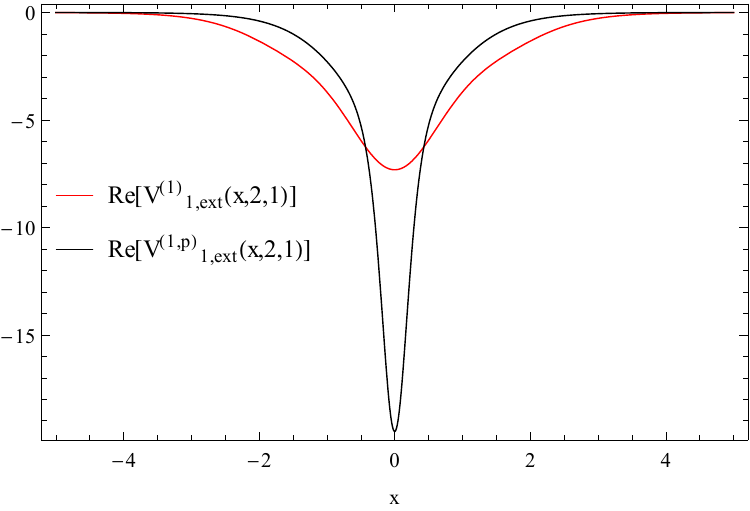}
\end{minipage}}
\hfill
 {\begin{minipage}[c][1\width]{0.45\textwidth}
\centering
 \includegraphics[scale=.52]{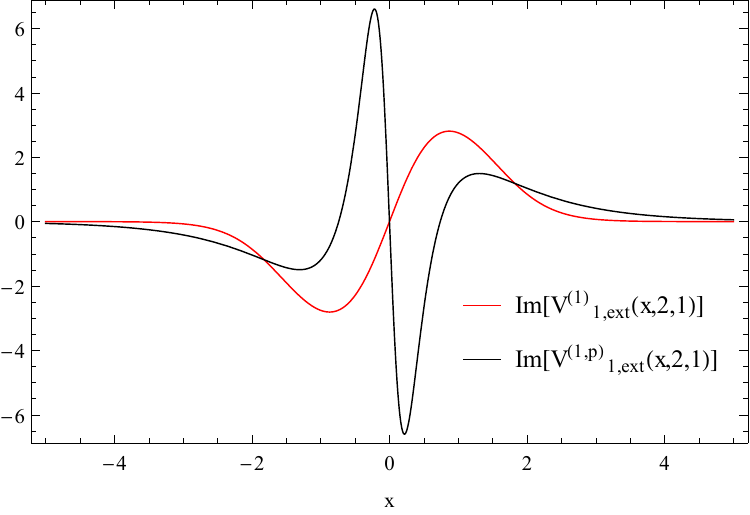}
\end{minipage}}
\hfill
 {\begin{minipage}[c][1\width]{0.45\textwidth}
\centering
 \includegraphics[scale=.52]{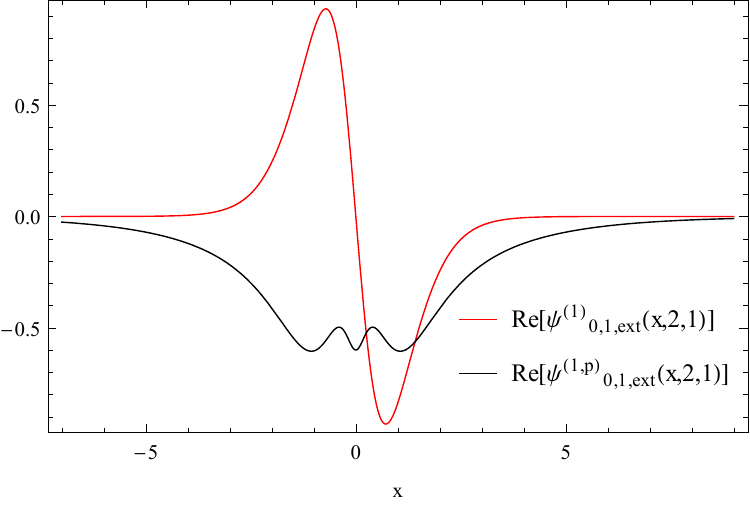}
 (a) $a=2, b=1 $
\end{minipage}}
\hfill
 {\begin{minipage}[c][1\width]{0.45\textwidth}
\centering
 \includegraphics[scale=.52]{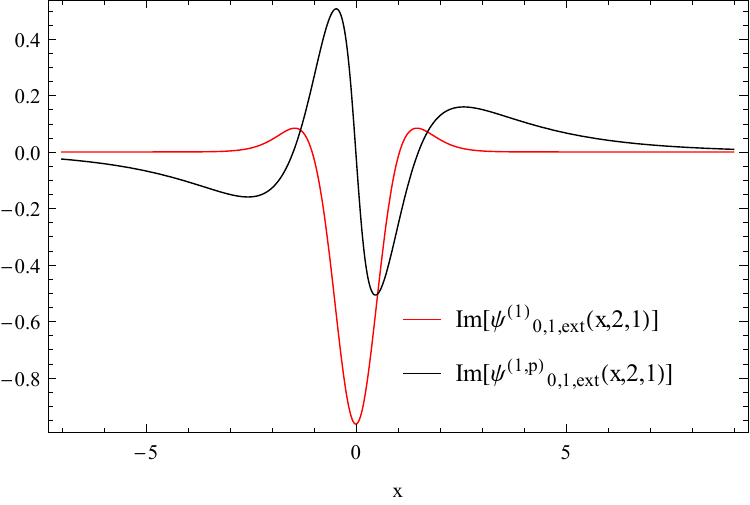}
 (b) $a=2, b=1 $
\end{minipage}}\\
{\bf Fig.8 (ii)}: {(a)-(b) Comparison between real and imaginary parts of RE $PT$ symmetric complex Scarf-II potential and their corresponding eigenfunctions obtained after parametric transformation  for integer combination of $a$ and $b$.}
\end{figure}


\section{Conclusions}
In this work, we have made an attempt to obtain all possible exactly solvable
complex PT-invariant reflectionless potentials. As a simple exercise,  
we first started with a well-known reflectionless real potential with $N$ bound
states and generated one continuous parameter ($\lambda$) family 
(which can be easily generalized to $N$-parameter family)
of strictly isospectral reflectionless potentials. As a special case we have 
also obtained expressions for the corresponding reflectionless Pursey and the 
AM potentials corresponding to $\lambda = 0$ and $-1$ respectively and with
$N-1$ bound states.
 
In the $PT$ symmetric case, we started with the well known complex 
PT-invariant Scarf-II potential and showed that it has novel parametric 
symmetry. We then showed that there are  $N$ number
of reflectionless potentials when both $a$ and $b$ are either integers or 
half-integers, thereby obtaining $2N$ number of complex PT-invariant 
reflectionless potentials in total. 
Further, we considered the rationally extended $PT$ symmetric reflectionless 
scarf-II potential, whose solutions are in terms of $X_m$-Jacobi EOPs
and shown that unlike the usual one, this extended potential is not 
invariant under the parametric symmetry but instead generates 
another set of reflectionless potentials whose solutions are also in terms of 
$X_m$-EOPs. By combining all these factors we then showed that there are
$2[(2N-1)m+N]$ number of complex PT-invariant reflectionless exactly solvable
potentials.

This paper raises few questions. Some of these are, have we really exhausted
the number of complex PT-invariant reflectionless exactly solvable 
potentials or are there are still more? While we believe that the answer to
the question is no, one can never be sure. Secondly, since reflectionless
potentials have found wide applications, it would be interesting if one of 
these complex reflectionless potential finds some application.

{\bf Acknowledgments}\\
AK is grateful to Indian National Science Academy (INSA) for awarding INSA Honorary
Scientist position at Savitribai Phule Pune University. BPM acknowledges the research
grant for faculty under IoE scheme (Number 6031) of Banaras Hindu University Varanasi.


\end{document}